\documentclass[epj]{svjour}

\usepackage[utf8]{inputenc}
\usepackage[T1]{fontenc}
\usepackage[english]{babel}
\usepackage[usenames,dvipsnames]{color}
\usepackage{graphicx}
\usepackage{amsmath}
\usepackage{amssymb}
\usepackage{hyperref}
\usepackage{textcomp}
\usepackage{etoolbox}
\usepackage{multirow}
\usepackage{url}
\usepackage{hyphenat}

\usepackage[numbers,square,sort&compress]{natbib}

\newcommand{\code}[1]{#1}

\newcommand{\inclxx}{\code{INCL++}}
\newcommand{\inclf}{\code{INCL4.6}}

\newcommand{\incl}{\code{INCL}}
\newcommand{\bic}{\code{BIC}}
\newcommand{\neutronhp}{\code{NeutronHP}}

\newcommand{\bertini}{\code{Bertini}}

\newcommand{\abla}{\code{ABLA07}}

\newcommand{\ablaold}{\code{ABLA~V3}}

\newcommand{\geh}{\code{G4ExcitationHandler}}
\newcommand{\ftfp}{\code{FTFP}}
\newcommand{\qgsp}{\code{QGSP}}
\newcommand{\ftfpinclxxhp}{\code{FTFP\_INCLXX\_HP}}
\newcommand{\qgspinclxxhp}{\code{QGSP\_INCLXX\_HP}}

\newcommand{\geant}{\code{Geant4}}

\newcommand{\highlightmissing}[1]{}
\newcommand{\ntof}{n\_TOF}

\ifpdf
  \graphicspath{{./figures/}}
\else
  \graphicspath{{./figures_eps/}}
\fi

\begin{document}

\title{On the role of secondary pions in spallation targets}

\author{
  Davide Mancusi\inst{1}\thanks{Corresponding author. E-mail address:
    davide.mancusi@cea.fr}
  \and
  Sergio {Lo Meo}\inst{2}\inst{3}
  \and
  Nicola Colonna\inst{4}
  \and
  Alain Boudard\inst{5}
  \and
  Miguel Antonio Cortés-Giraldo\inst{6}
  \and
  Joseph Cugnon\inst{7}
  \and
  {Jean-Christophe} David\inst{5}
  \and
  Sylvie Leray\inst{5}
  \and
  Jorge Lerendegui-Marco\inst{6}
  \and
  Cristian Massimi\inst{3}\inst{8}
  \and
  Vasilis Vlachoudis\inst{9}
}
\institute{Den-Service d'étude des réacteurs et de mathématiques appliquées
  (SERMA), CEA, Université Paris-Saclay, F-91191, Gif-sur-Yvette, France
  \and
  ENEA, Research Centre ``Ezio Clementel'', I-40129 Bologna, Italy
  \and
  INFN, Section of Bologna, I-40127 Bologna, Italy
  \and
  INFN, Section of Bari, I-70125 Bari, Italy
  \and
  IRFU, CEA, Université Paris-Saclay, F-91191, Gif-sur-Yvette,
  France
  \and
  Universidad de Sevilla, Facultad de Fisica, 41012 Sevilla, Spain
  \and
  AGO department, University of Li\`{e}ge, all\'{e}e du 6 ao\^{u}t
  17, b\^{a}t.~B5, B-4000 Li\`{e}ge 1, Belgium
  \and
  Physics and Astronomy Dept. ``Alma Mater Studiorum'' - University
  of Bologna, I-40126 Bologna, Italy
  \and
  European Organization for Nuclear Research (CERN), CH-1211 Geneva,
  Switzerland
}

\date{Received: \today}

\abstract{
  We use particle-transport simulations to show that secondary pions play a
  crucial role for the development of the hadronic cascade and therefore for the
  production of neutrons and photons from thick spallation targets. In
  particular, for the \ntof{} lead spallation target, irradiated with
  $20$~GeV${}/c$ protons, neutral pions are involved in the production of
  $\sim90$\% of the high-energy photons; charged pions participate in $\sim40$\%
  of the integral neutron yield. Nevertheless, photon and neutron yields are
  shown to be relatively insensitive to large changes of the average pion
  multiplicity in the individual spallation reactions. We characterize this
  robustness as a peculiar property of hadronic cascades in thick targets.
}

\PACS{
  {25.40.Sc}{Spallation reactions} \and
  {24.10.Lx}{Monte-Carlo simulations} \and
  {28.20.Gd}{Neutron transport: diffusion and moderation}
}


\maketitle

\section{Introduction}
\label{sec:introduction}

In spallation reactions, a high-energy (${}>150$~MeV) light projectile collides
with a nucleus and on average leads to the emission of a large number of
particles, mostly neutrons. The spectrum of spallation neutrons extends to large
energies, up to the energy of the incoming projectile. For this reason,
spallation reactions are often used for the purpose of generating intense
high-energy neutron fluxes \cite{filges-handbook}, as it is the case for
instance in Accelerator-Driven Systems (ADS), subcritical reactor cores that are
kept in a steady state by neutrons produced by a spallation source
\cite{ait_abderrahim-myrrha}.

Neutrons are not the only particles that are emitted during spallation
reactions. Protons and light charged particles (LCPs, $A\leq4$) are also
present, as are pions if the projectile energy is high enough. Spallation is
actually capable of producing (with varying yields) all nuclei lighter than the
target nucleus and close to the stability valley, as well as a handful of nuclei
heavier than the target nucleus (as amply demonstrated by several experimental
campaigns \cite[see e.g.][Fig.~12]{enqvist-lead}). All these particles,
especially the lightest ones (neutrons, protons, pions and LCPs), are capable of
inducing secondary nuclear reactions in a thick spallation target, and may thus
contribute to the development of the hadronic cascade, to particle emission and
to the production of residual nuclei \cite[see e.g.][]{david-At}.

The standard theoretical tool for the description of spallation reactions is a
hybrid nuclear-reaction model where an intranuclear-cascade (INC) stage is
followed by an optional pre-equilibrium stage and by a statistical de-excitation
stage \cite{filges-handbook}. For the reasons evoked in the previous paragraph,
these models must be validated not only for the primary reactions (typically
reaction between fast protons and heavy nuclei such as tungsten, lead or
bismuth), but also for all secondary reactions that may sizably contribute to
the production of neutrons or to any other observable one may be interested
in. It is generally acknowledged that secondary proton- and neutron-nucleus
reactions are important, as suggested by the selection of validation data for
international nuclear-reaction-model intercomparisons
\cite{david-intercomparison,leray-intercomparison,intercomparison-website};
however, the same intercomparisons devoted little attention to the production of
secondary pions and to the validation of models on pion-induced reactions. This
is at least partly due to the fact that inclusive data for pion-nucleus
reactions are scarce, and partly to the fact that ADSs are expected to operate
at energies of the order of $1$~GeV \cite{ait_abderrahim-myrrha}, where pion
multiplicities are relatively low.

Several spallation neutron sources are currently operational around the world
and more are under construction or planned for the near future. Predictions of
the neutron-source characteristics can typically be obtained by means of Monte
Carlo (MC) simulations. Reliable results require detailed and accurate knowledge
of the physical processes at the basis of the spallation reactions. Among the
currently operating spallation neutron sources, the \ntof{} (neutron
Time-Of-Flight) facility \cite{guerrero-ntof} is an intense pulsed neutron
source located at CERN. Neutrons are produced by spallation of lead nuclei
caused by an incident $20$~GeV${}/c$ proton beam, and subsequently moderated and
collimated towards two experimental areas, where their energy can be measured
using the time-of-flight technique. One of the foremost advantages of the
detection capability of \ntof{} is that the produced neutrons extend over more
than twelve orders of magnitude, from thermal energies to the GeV range,
allowing highly accurate measurements for a wide range of applications. Precise
characterization of the neutron source is crucial for these purposes, and some
features of the neutron beam must be inevitably determined via numerical
simulations \cite{ntof-fluka}. In recent publications
\cite{lo_meo-ntof_g4,lerendegui-ntof_ear2}, the \geant{} toolkit for particle
transport \cite{agostinelli-geant,allison-geant} was used to characterize the
neutron and photon fluxes directed towards the \ntof{} experimental
areas. Calculations of neutron and photon fluences performed with different
\geant{} physics lists exhibited large relative differences. The authors
suggested at the time that this difference could be related to different
treatments of pion production and pion-induced reactions.

In this work, we study the role of pion production and its influence on the
spallation yields. In particular, it will be shown that secondary pions play a
crucial role for particle production in thick spallation targets, such as the
\ntof{} neutron source. We shall demonstrate that the production of high-energy
prompt photons is essentially dominated by $\pi^0$ decay; this phenomenon is
well known in the context of the phenomenology of calorimetric measurements for
high-energy physics \cite{fabjan-calorimetry_review}. At the same time, the
production of neutrons is affected by both secondary $\pi^\pm$-nucleus reactions
and $\pi^0$ production.  These facts notwithstanding, particle yields are less
sensitive to the detail of the specific nuclear-reaction model used for the
particle-transport simulation. This is explained in terms of an intrinsic
``resilience'' of hadronic cascades in thick targets.

The paper is structured as follows. In Sec.~\ref{sec:model-descr-liege} we
provide a brief description of the salient features of the Liège
Intranuclear-Cascade model (\incl{}), which is pivotal for our numerical
simulations of the \ntof{} spallation target.  Thin-target model calculations
related to the pion sector are presented and discussed in
Sec.~\ref{sec:thin-target:-pion}, along with comparisons against experimental
data. Section~\ref{sec:thick-targ-second} shifts the focus towards the
thick-target transport calculations of the \ntof{} spallation target. The most
important features of the MC simulations, described in detail in recent papers
\cite{lo_meo-ntof_g4,lerendegui-ntof_ear2}, are recalled in
Secs.~\ref{sec:geant-toolkit} and \ref{sec:ntof-simulation}. The role played by
pions in the emission of neutrons and photons is highlighted in
Secs.~\ref{sec:analys-second-react}--\ref{sec:infl-charg-pions}. Section~\ref{sec:infl-pion-mult}
illustrates the tendency of the hadronic cascade to mitigate the sensitivity of
the particle yields to the details of the description of the nuclear
reactions. Conclusions are drawn in Sec.~\ref{sec:conclusions}.

\section{Model description: the Liège Intranuclear Cascade model}
\label{sec:model-descr-liege}

The Li\`{e}ge Intranuclear Cascade model (\incl)
\cite{boudard-incl,mancusi-inclxx} is one of the most refined existing tools for
the description of spallation reactions. The model is currently maintained and
developed jointly by the University of Li\`{e}ge (Belgium) and CEA (Saclay,
France).  The model assumes that the first stage of the reaction can be
described as an avalanche of independent binary collisions.  The \incl{} model
is essentially classical, with the addition of a few suitable ingredients that
mimic genuine quantum-mechanical features of the initial conditions and of the
dynamics: for instance, target nucleons are endowed with Fermi motion, realistic
space densities are used, the output of binary collisions is random and
elementary nucleon-nucleon collisions are subject to Pauli blocking.  The model
can describe the emission of nucleons and pions; light clusters (up to $Z=5$,
$A=8$ by default) can also be produced through a dynamical phase-space
coalescence algorithm. 

Intranuclear-cascade models in general (and \incl{} in particular) only describe
the fast, dynamical stage of a spallation reaction, leading to the formation of
excited nuclei which subsequently de-excite by emitting particles and/or
fissioning. It is therefore necessary to follow the de-excitation of this
cascade \emph{remnant} if one requires a complete description of the nuclear
reaction.  Since the time scale for de-excitation is much longer than for
cascade, a different physical description is usually employed. This may include
an optional pre-equilibrium stage, which then handles the thermalization of the
remnant; if pre-equilibrium is used, the intranuclear-cascade stage is stopped
earlier. Either way, thermalization is attained and subsequent de-excitation of
the remnant is described by statistical de-excitation models. Within \geant{},
\incl{} can be directly coupled with two different de-excitation codes, namely:
\geh{}, the native statistical de-excitation model of \geant{} and the default
choice \cite{quesada-g4excitationhandler}, and \ablaold{}, a de-excitation model
developed at GSI (Darmstadt, Germany) \cite{gaimard-abla,junghans-abla}. We
stress here that this is not the code that is usually coupled to \incl{} (which
is \abla{} \cite{kelic-abla07}), but rather an older version. A detailed
comparison of the capabilities of the two versions can be found in
Ref.~\citenum{kelic-abla07}.

Different particles are produced in different stages of the spallation
reactions. In particular, while neutrons and $\gamma$-rays are mostly generated
in de-excitation processes, pion production, in particular, occurs entirely in
the first reaction stage. The pion dynamics in \incl{} has been recently
upgraded to push the upper energy limit of the model up to 15--20~GeV. Older
versions of \incl{} considered only one mechanism for pion production, namely
excitation and subsequent decay of the $\Delta(1232)$ resonance. For
nucleon-induced reactions, this is a good approximation up to energies of about
$2$--$3$~GeV. This is proven by the results of the IAEA benchmark
\cite{david-intercomparison,leray-intercomparison,intercomparison-website}, as
well by the previous studies on the \incl{} pion dynamics
\cite{aoust-pion_physics,aoust-potential}. Additionally, one should not forget
that, as soon as multiple collisions are involved, any particle correlation due
to the action of an intermediate resonance will be washed out. For the purpose
of correctly describing multiple-collision reactions, it is sufficient to
capture the first-order behavior, and correlations may be neglected. Of course,
selective or exclusive observables (such as two-particle correlations),
especially if related to one- or few-collision reactions will generally be
incorrectly described.

Above $2$--$3$~GeV, excitation of heavier baryonic and mesonic resonances
becomes likely\footnote{The excitation of the Roper $N^*(1440)$ resonance is a
  special exception, because it may be excited at lower energy in the $T=0$
  channel. \inclf{} assumes that the kinematics of pion production in this
  channel is governed by the $\Delta(1232)$ resonance. The extended version of
  \incl{} does not make this approximation.}. A straightforward extension of INC
would in principle entail the description of all the energy-angle-differential
cross sections for the formation, scattering and absorption of the resonances in
the nuclear medium, as well as their mean-field potentials, decay modes,
etc. The amount of information that must be fed into the model is ponderous;
besides, most of the time, the available experimental information on these
elementary processes is direly scarce, or partial at best. One possible approach
would be to rely on an independent event generator for the elementary
hadron-nucleus collisions, in the spirit of Ref.~\citenum{ackerstaff-MICRES}. In
this paper, however, we explore a different solution.

It should be noted that baryonic resonances above $\Delta(1232)$ are largely
overlapping. This raises the question of whether it is meaningful to consider
them as having separate identities in the framework of INC. Additionally, their
lifetime (in vacuum) is much smaller than the typical time between subsequent
collisions during INC (a few fm${}/c$), so that a heavy resonance is unlikely to
undergo any collision before decaying in the nucleus. This is already marginally
the case for $\Delta(1232)$, whose lifetime in vacuum is $\sim1.6$~fm${}/c$, and
indeed most of the observables calculated in INC are rather insensitive to
variations of the $\Delta(1232)$ lifetime. It should also be considered that the
final (on the time scale of INC) decay products of baryonic resonances are often
pions.

Strictly speaking, the arguments above do not apply to most of the lightest
unflavored mesonic resonances ($\eta$, $\omega$, $\eta'$\ldots), whose lifetimes
are comparable to or longer than the duration of the INC stage; nor do they
apply to strange baryons and mesons ($\Lambda$, $\Sigma$, $K$\ldots), which
undergo weak decay. However, the available experimental elementary cross
sections associated with the production of these particles
\cite{flaminio-hera_1,flaminio-hera_3} and order-of-magnitude estimates suggest
that their global influence on the INC dynamics is weak. Therefore, it should be
possible to treat them as corrections, at least in the energy range up to
$10$--$15$~GeV.

In view of the discussion above, it is appropriate to use a more pragmatic
approach to the description of high-energy reactions. In the latest version of
the \incl{} model, the production and decay of individual resonances (except for
$\Delta(1232)$) is bypassed and replaced by \emph{multipion collisions}, i.e.\
effective two-body collisions leading to the production of one or more pions in
the final state, of the following form:
\begin{subequations}
  \label{eq:multipion_channels}
  \begin{align}
    N+N&{}\rightarrow N+N+x\pi\text,\\
    \pi+N&{}\rightarrow N+x\pi\text.
  \end{align}
\end{subequations}
In the current model, the number $x$ of pions in the final state of the
collision takes all values from $1$ to $4$ inclusive.

The rest of the pion dynamics in the new version of \incl{} is the same as in
the older one. The formation, absorption and decay of the $\Delta(1232)$
resonance is explicitly treated. Pion absorption is possible only via the
formation of $\Delta(1232)$. No one-step mechanism for pion absorption on
nucleon pairs is included.  Further details on the latest version of \incl{} can
be found in Refs.~\citenum{pedoux-pions,pedoux-phd}.  Relative to the published
version, the current implementation of the model has slightly evolved, with the
most notable difference concerning the biasing of nucleons towards the forward
direction in the center-of-mass system\footnote{In the current model, the
  final-state particle momenta are generated according to a flat, unbiased
  phase-space sampling algorithm. Let $E$ be the generated CM energy of the
  first nucleon; the value of $E$ determines the minimum ($t_\text{min}$) and
  maximum ($t_\text{max}$) values of the Mandelstam four-momentum transfer
  $t$. The value of $t$ is then sampled from a distribution of the form
  $\exp{(Bt)}$ and \emph{all} the generated momenta are rotated to match the
  sampled four-momentum transfer for the first nucleon. Clearly this algorithm
  does not modify the single-particle energy distributions in the CM system,
  which are therefore still given by the phase-space model. On the other hand,
  the distributions in the laboratory system are different.}.

\begin{figure}
  \centering
  \includegraphics[width=\linewidth]{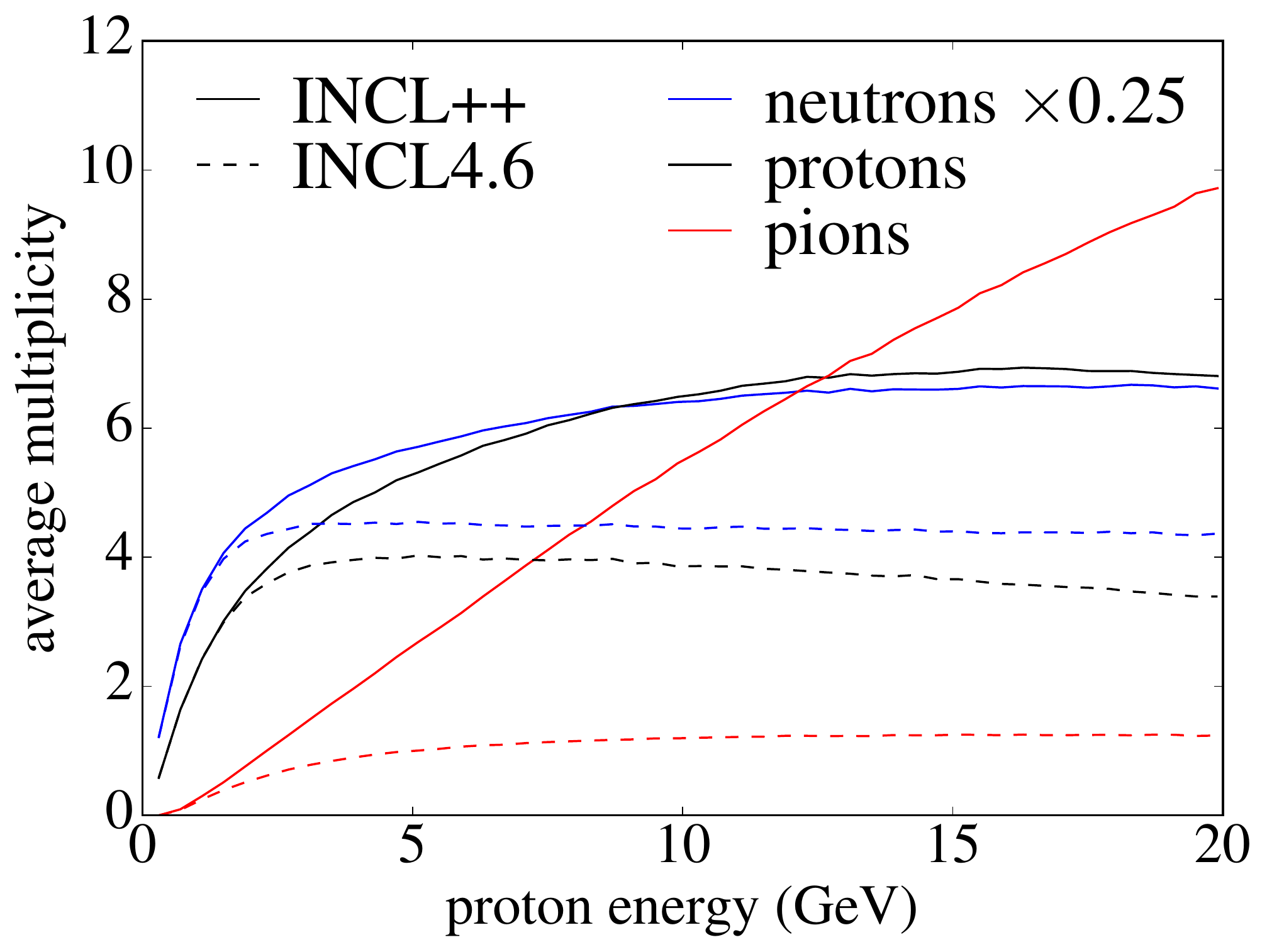}
  \caption{Excitation function for the total average neutron, proton and pion
    multiplicities (from intranuclear cascade and de-excitation) in the final
    state of p+$^{208}$Pb reactions, as calculated with (\inclxx{}) and without
    (\inclf{}) multipion extension, coupled with \abla{}. Note that the neutron
    curve has been renormalized by a factor of $0.25$.}
  \label{fig:multiplicities}
\end{figure}

\begin{figure}
  \centering
  \includegraphics[width=\linewidth]{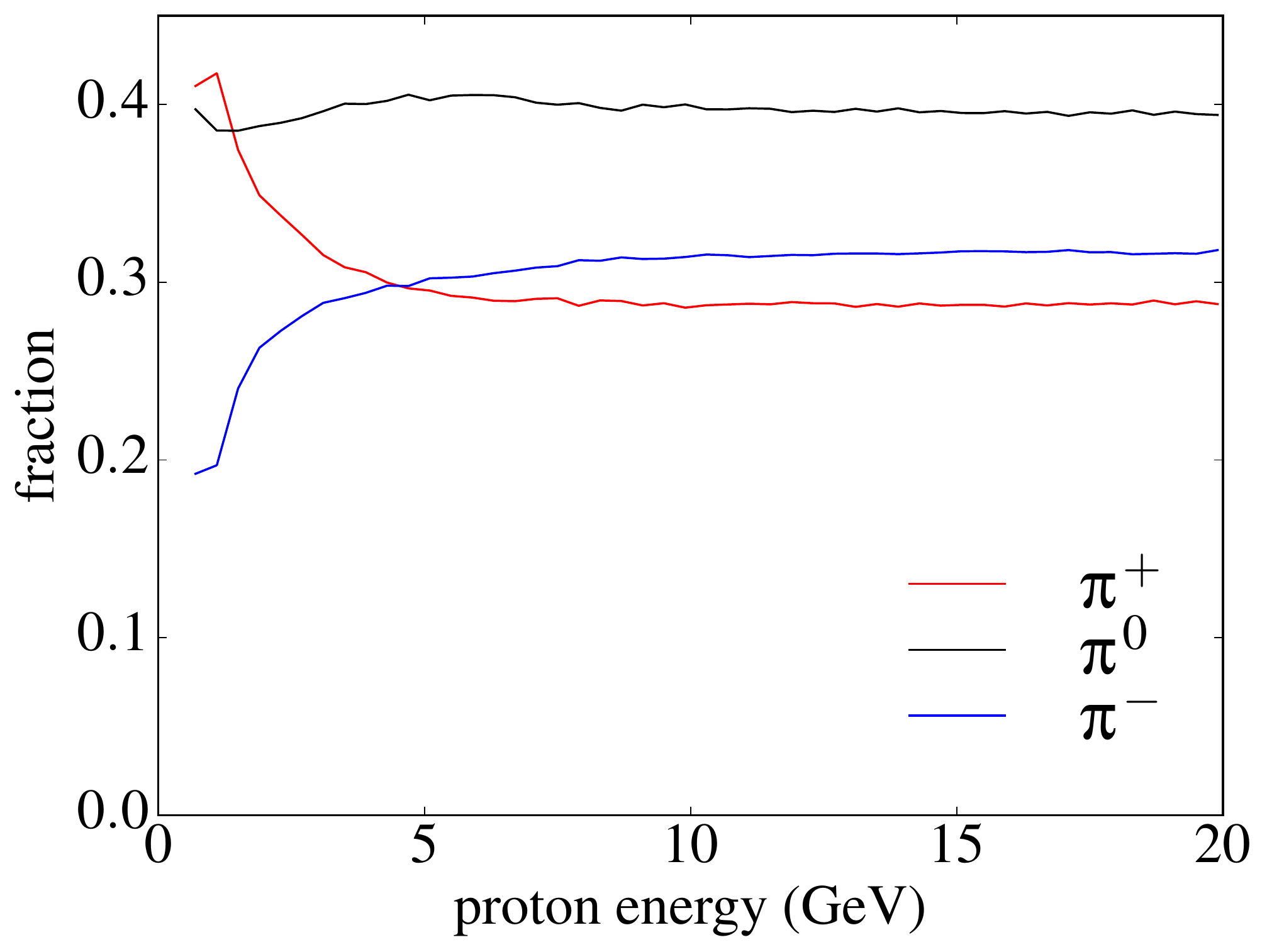}
  \caption{Excitation functions for the average fraction of produced pions for
    each of the three charge states, in p+$^{208}$Pb, as calculated by
    \inclxx{}.}
  \label{fig:pion_charge_repartition}
\end{figure}

The effect of the multipion extension can be studied by comparing some global
quantities calculated with the old (\inclf{}) and the new extended model
(\inclxx{}). Figure~\ref{fig:multiplicities} shows the average pion multiplicity
(i.e.\ the average number of pions produced per inelastic reaction) in
p+$^{208}$Pb as a function of the proton energy. While the two models yield
similar predictions at low projectile energy, in the older model the
multiplicity saturates around $5$~GeV, never exceeding $\sim1$~pion per
reaction, while the extended model yields an almost linear increase up to
$20$~GeV. Figure~\ref{fig:pion_charge_repartition} shows how the produced pions
are distributed over the three charge states, according to the calculations of
the extended model. The fraction of neutral pions is roughly
energy-independent. On the contrary, the lines for positive and negative pions
cross between $4$ and $5$~GeV. The suppression of negative pions at low energy
can be explained by considering that the projectile (a proton) carries positive
isospin, and that pions can only be produced in the first few collisions. As the
energy and the number of collisions increase, pion production becomes
increasingly dominated by the total isospin of the system, which is negative
because $N>Z$ in lead. Therefore, negative pions are asymptotically more
abundantly produced than positive pions.

For completeness, we mention that the version of the \incl{} model that was used
for the present work is \inclxx{} \code{v5.2.9.2}.

\section{Thin target: pion-production cross
  sections}\label{sec:thin-target:-pion}

As discussed in Sec.~\ref{sec:introduction}, \geant{} simulations performed with
an \inclxx{}-based physics list yield the best overall reproduction of the
measured neutron production for the \ntof{} spallation target, contrary to the
physics lists using the Binary Cascade (BIC) \cite{folger-g4bic} or Bertini
models \cite{wright-bertini}, which overestimate the experimentally evaluated
neutron production by as much as 70\% \cite{lo_meo-ntof_g4}. In
Ref.~\citenum{lo_meo-ntof_g4} it was hinted that a possible explanation of this
difference could be related to pion production. In particular, it was pointed
out that both neutral and charged pions could play an important role in
determining the production of neutrons as well as of the so-called prompt
$\gamma$-ray component, i.e. those produced in the first nanoseconds of the
spallation reactions (with the delayed $\gamma$-ray component produced later on
from neutron capture reaction and de-excitation of excited residues). In the
following, the role of secondary pions in spallation targets is investigated,
starting from a comparison of theoretical differential cross sections with the
available experimental data. We remark that the predictive capability of the
\incl{} model for the production of other particles (in particular neutrons,
protons and light charged particles) below $3$~GeV has already been established
in an extensive benchmark of spallation models, organized under the auspices of
the IAEA
\cite{david-intercomparison,leray-intercomparison,intercomparison-website}.

In order to assess the validity of the \inclxx{} and other models, it is very
useful to compare with one of the most complete and comprehensive data set on
pion production at high energy. Such data were collected by the HARP experiment
at CERN \cite{catanesi-harp,apollonio-harp_pion_beams}, where extensive
measurements of double-differential cross sections for charged-pion production
in proton- and pion-induced reactions were performed. Incident momenta of $3$,
$5$, $8$ and $12$~GeV${}/c$ were considered.

\subsection{Integral pion production}

\begin{figure}
  \centering
  \includegraphics[width=\linewidth]{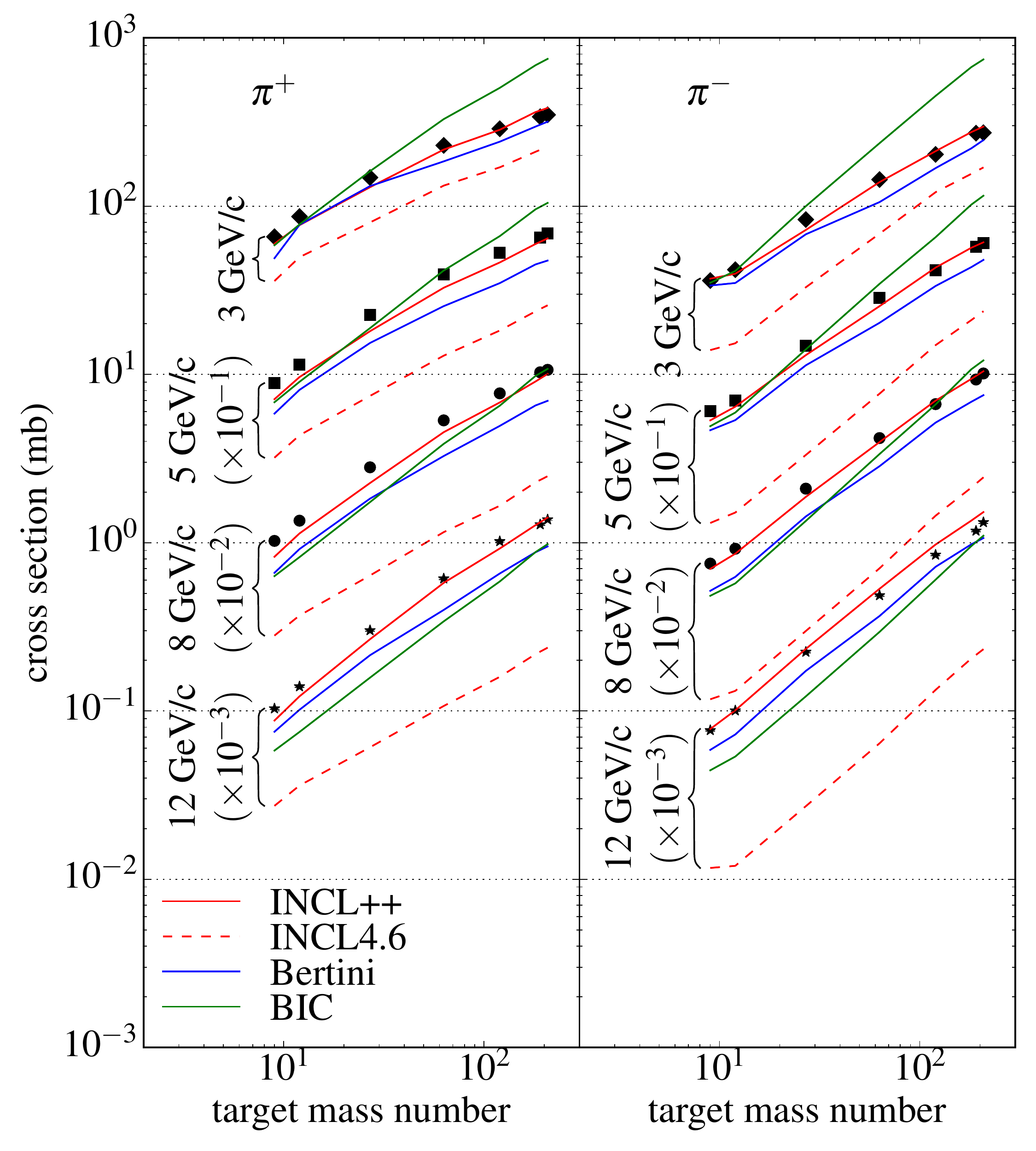}
  \caption{Cross sections for the production of $\pi^+$ (left) and $\pi^-$
    (right) from proton-nucleus reactions, integrated over the HARP
    angle-momentum acceptance, for different incident proton momenta, as
    functions of the target mass number. The lines represent calculations by
    different models (see text for details).  The experimental data are taken
    from Ref.~\citenum{catanesi-harp}.}
  \label{fig:p_pi_yields}
\end{figure}

\begin{figure}
  \centering
  \includegraphics[width=\linewidth]{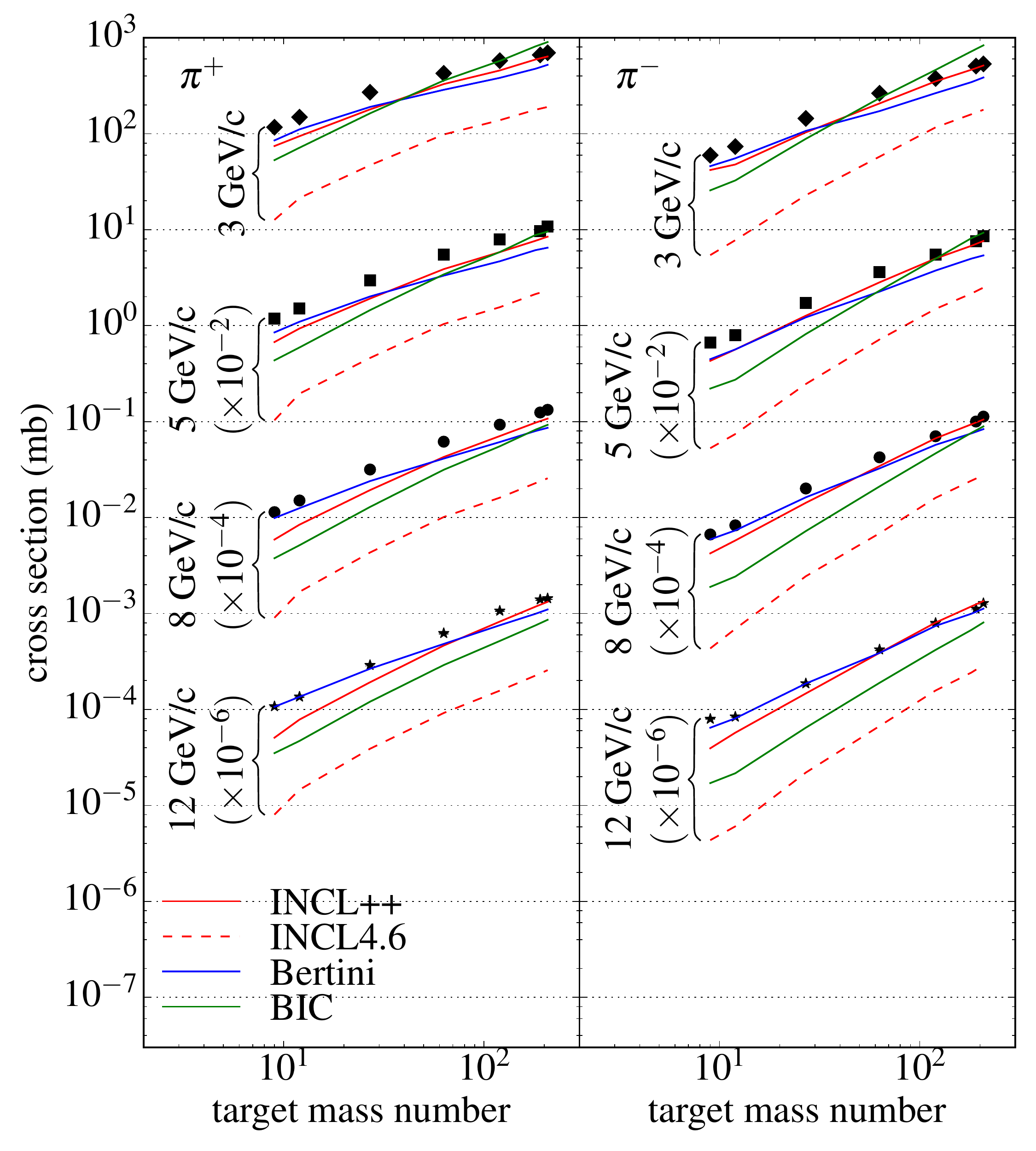}
  \caption{Same as Fig.~\ref{fig:p_pi_yields}, but for $\pi^+$-nucleus
    reactions.}
  \label{fig:pip_pi_yields}
\end{figure}

\begin{figure}
  \centering
  \includegraphics[width=\linewidth]{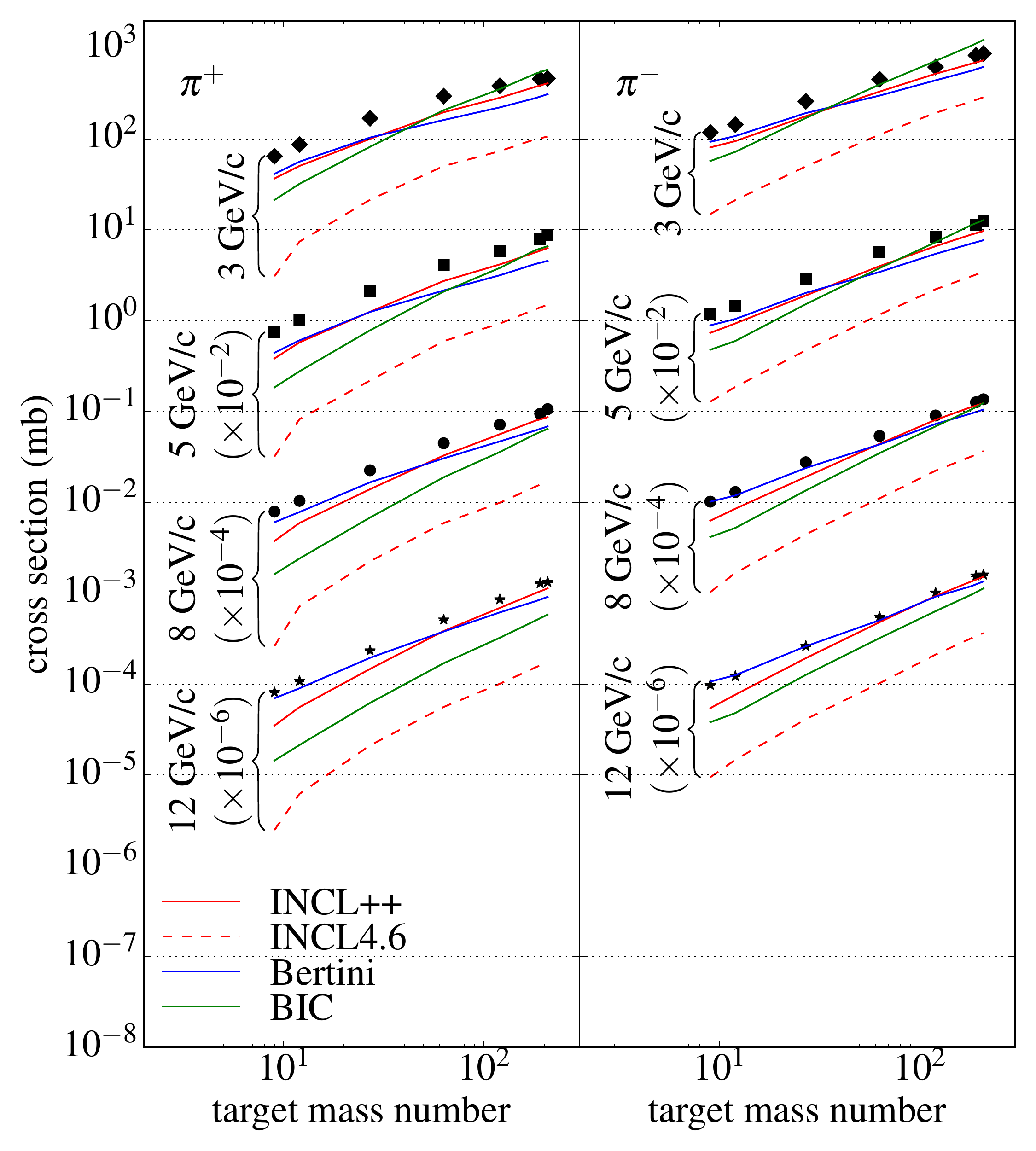}
  \caption{Same as Fig.~\ref{fig:p_pi_yields}, but for $\pi^-$-nucleus
    reactions.}
  \label{fig:pim_pi_yields}
\end{figure}

Figures~\ref{fig:p_pi_yields}--\ref{fig:pim_pi_yields} show inclusive
pion-production cross sections integrated over the acceptance of the HARP
experiment. In addition to the \inclxx{} calculation, we show the results of
three other models: \bertini{} \cite{wright-bertini} and Binary Cascade (\bic{})
\cite{folger-g4bic} are popular intranuclear-cascade models available in
\geant{}, while \inclf{} represents the Liège Intranuclear Cascade model
\emph{without} multipion extension \cite{boudard-incl4.6}.  The difference
between \inclf{} and \inclxx{} clearly highlights the importance of the
extension, which is already sizable at the lowest incident momentum of the HARP
data-set ($3$~GeV${}/c$). The \inclf{} model is reported to illustrate certain
surprising features of the hadronic shower in Sec.~\ref{sec:thick-targ-second},
namely the relative insensitiveness to the details of the treatment of the
individual elementary interactions.

The \inclxx{} and \bertini{} models provide comparably accurate
predictions. \bertini{} is generally closer to the experimental data for light
targets, while \inclxx{} performs better on heavy targets, such as lead, which
is most interesting for the present work and in general for spallation neutron
sources.  Proton-nucleus data are generally better reproduced than pion-nucleus
data, with all calculations anyway being within a factor of 2 from the
experimental data, with the \bic{} model having greater difficulties in
reproducing the experimental cross sections.

It is important to remark that very few inclusive experimental data exist for
the production of neutral pions in proton-nucleus and pion-nucleus
reactions. This is of course mainly due to the short lifetime of the neutral
pion, which complicates its detection. It is therefore customary to benchmark
reaction models only on charged-pion production. We will follow the same
approach in the present work. The validity of the interpolation to neutral pions
can often be directly related to the goodness of the isospin-symmetry
approximation, which is commonly used for the computation of elementary cross
sections in intranuclear cascade.

\subsection{Double-differential pion-production cross sections}





\begin{figure*}
  \centering
  \includegraphics[width=\linewidth]{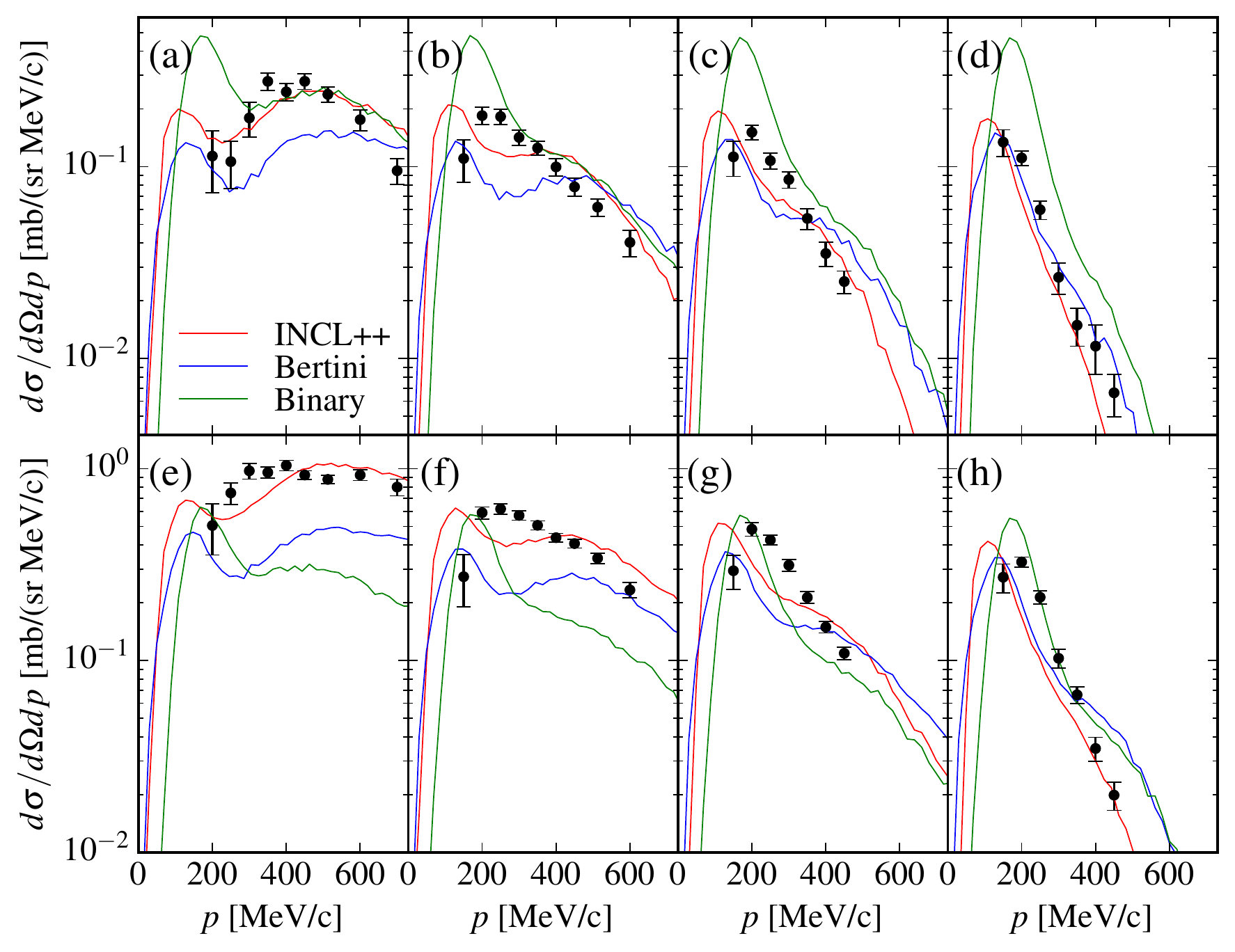}
  \caption{Double-differential cross sections for the production of $\pi^+$ at
    $25^\circ$ (a, e), $48^\circ$ (b, f), $71^\circ$ (c, g) and $105^\circ$ (d,
    h), from 3~GeV${}/c$ (a--d) and 12~GeV${}/c$ (e--h) p+Pb. The lines
    represent calculations by different models (see text for details). The
    experimental data are taken from Ref.~\citenum{catanesi-harp}.}
  \label{fig:p_Pb_compact_ddxs}
\end{figure*}

\begin{figure*}
  \centering
  \includegraphics[width=\linewidth]{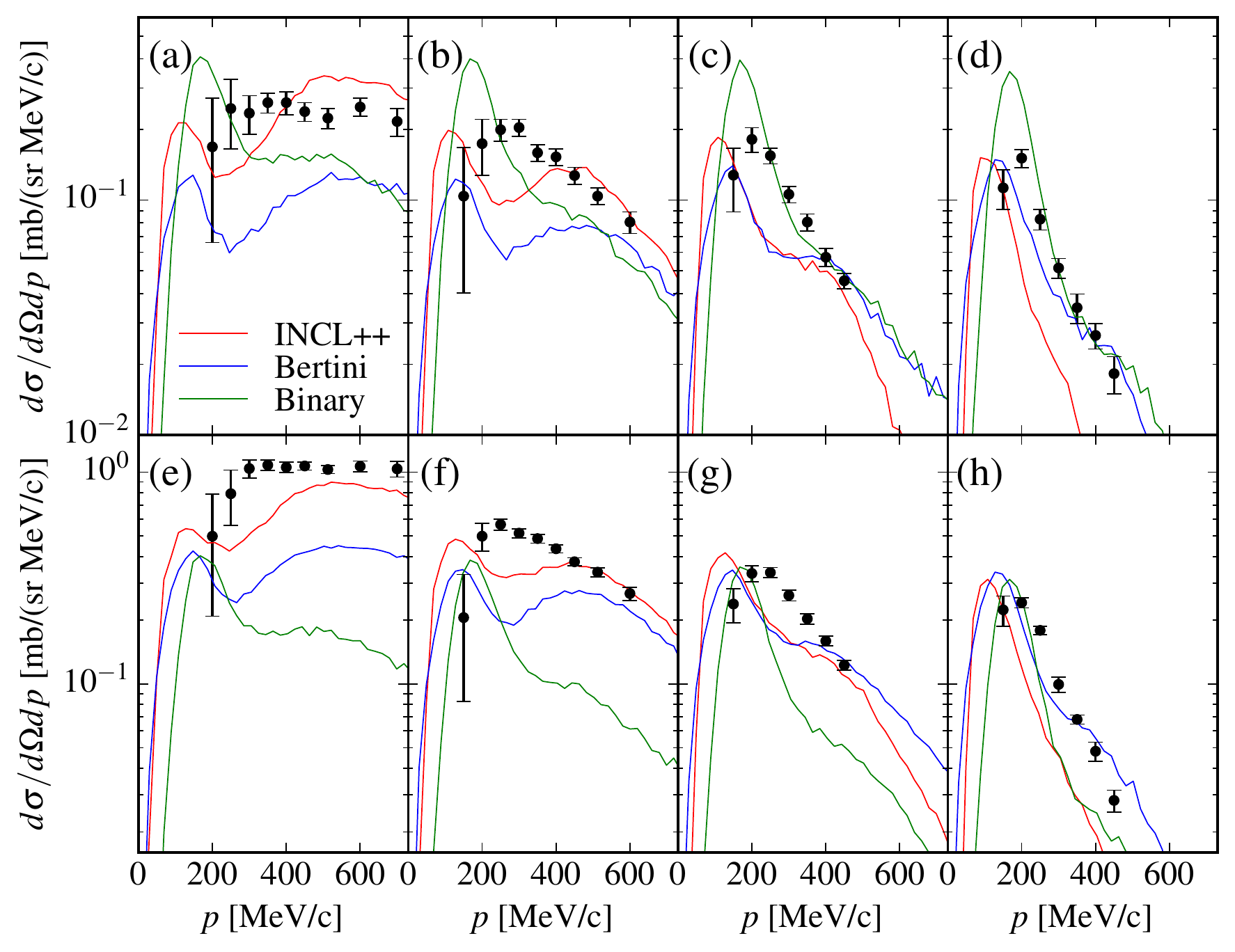}
  \caption{Same as Fig.~\ref{fig:p_Pb_compact_ddxs}, but for 3~GeV${}/c$ (a--d)
    and 12~GeV${}/c$ (e--h) $\pi^-$+Pb reactions.}
  \label{fig:pim_Pb_compact_ddxs}
\end{figure*}

Figures~\ref{fig:p_Pb_compact_ddxs} and \ref{fig:pim_Pb_compact_ddxs} show
double-differential (momentum-angle) cross sections for inclusive pion
production in proton- and pion-induced reactions. For benchmarking we select two
incident momenta - $3$ and $12$~GeV${}/c$ - and we focus on the lead target,
which is the most important for the study of the \ntof{} spallation source. For
simplicity, we limit our discussion to $\pi^+$ production in proton- and
$\pi^-$-induced reactions; these results exhibit all the typical features of the
general case.

For the purpose of this work, the most interesting quantity to compare is the
pion emission spectrum. The cross sections of
Figs.~\ref{fig:p_pi_yields}--\ref{fig:pim_pi_yields} are determined by
integration of the double-differential cross sections in
Figs.~\ref{fig:p_Pb_compact_ddxs} and \ref{fig:pim_Pb_compact_ddxs} over the
momentum and angle acceptance of the HARP data-set. It clearly appears that no
model accurately reproduces the emission spectra for all angles and
momenta. \inclxx{} and \bertini{} are generally more accurate at forward and
backward angles, respectively, while \bic{} is, as already noted, rather far
from the experimental data. The goodness of the model predictions for this
observable is qualitatively consistent with the results for neutron production
in \geant{} simulations of the \ntof{} spallation target, providing further
evidence of the fundamental role of pion-induced reactions in thick spallation
targets.

An interesting observation that can be made about
Figs.~\ref{fig:p_Pb_compact_ddxs} and \ref{fig:pim_Pb_compact_ddxs} is that
\inclxx{} and \bertini{} consistently show a dip in the spectra at forward
angles (even up to roughly $90^\circ$) and around $250$~MeV${}/c$, which is not
seen in the experimental data. This defect was also noticed by the authors of
Ref.~\citenum{wright-bertini}, who tentatively attributed it to insufficient
moderation by the nuclear medium. In our opinion, however, the dip is related to
the formation and decay of the $\Delta(1232)$ resonance, which manifests itself
as a strong peak in the pion-nucleon cross section. This intuition is triggered
by the observation that the position of the dip coincides approximately with the
position of the resonance in the $\pi+N\to\Delta$ cross section. Indeed, we have
verified that the dip is insensitive to reasonable modifications of the
recombination ($\Delta+N\to N+N$) cross section.

The mechanism leading to the formation of the dip in the model is rather simple,
if one makes a few reasonable assumptions. First, we assume that pion production
in INC proceeds in two stages. In the first stage, early elementary collisions
generate a structureless (no dip) pion spectrum (it is reasonable to assume that
pions are produced early in the reaction because the energy available for pion
production quickly degrades after a few collisions). In the second stage, the
generated pions traverse the nucleus, possibly undergoing scattering and
absorption, and possibly emerging as free particles. In this picture, the early
pions are attenuated by the nuclear medium, with the excitation of the
$\Delta(1232)$ resonance playing a role in the distortion of the pristine pion
spectrum, due to selective pion absorption at the corresponding resonance
energy.  Since the dip is insensitive to the recombination cross section and to
the resonance lifetime, and since $\Delta$ resonances (in \incl{}) can only be
absorbed by recombination, we conclude that the intermediate $\Delta$ resonances
mostly decay back to pion-nucleon pairs. In principle, the momentum of the pion
should fall back in the dip region. However, while the formation and decay of
the intermediate $\Delta$ resonances does not modify the pion momentum
distribution, it does act on the angular distribution. If one makes the
reasonable assumption that the pristine pion spectrum is sensibly
forward-peaked, then the decay of the intermediate $\Delta$ resonances will
redistribute pions from the forward angles to all angles. This manifests itself
as a dip at forward angles in the double-differential spectra.

While this explanation might hold valid for the dip observed in the model
calculations, it is not clear whether it also applies to the data. A hint of a
dip may be seen in the very forward angles, but in general data seems to
indicate that in reality the dip, if any, is less pronounced than what predicted
by the models.  We performed some tests and we verified that the dip disappears
if the $\pi+N\to\Delta$ cross section is artificially reduced by about a factor
of $2$. One can also act on the width of the $\Delta$ resonance peak in the
$\pi+N\to\Delta$ entrance channel: interestingly, either increasing or
decreasing the width of the Breit-Wigner-like peak will suppress the dip in the
calculations. Theoretical calculations \cite[e.g.][]{oset-delta_self_energy}
indicate that in-medium $\Delta$ resonances should be broader than the
corresponding free particles, although unambiguous quantitative indications are
still missing \cite{ter_haar-deltas}. \incl{} already generates part of this
medium effect (on the resonance lifetime) through the application of Pauli
blocking on the resonance decay and through $\Delta$ absorption. For consistency
one should also modify the cross section of the formation process to reflect
this. It remains to be seen if a realistic modification of the $\Delta$ width
(in the spirit of e.g.\ Refs.~\citenum{yariv-isabel1,yariv-isabel2}) can
reconcile the calculations with the experimental data.

For the sake of completeness, we mention that there is disagreement about the
scientific adequateness of the HARP data analysis.  A group of former HARP
collaborators (the HARP-CDP group) have published a revisited analysis of the
raw HARP data, which has sparked a long and well-documented controversy
\cite{harp-cdp-website}. Ref.~\citenum{bolshakova-harp_cdp_lead} contains direct
comparisons of the double-differential momentum-angle cross sections, but only
for the smallest angle ($25^\circ$) and for $3$ and $8$~GeV${}/c$ beam momenta
(Figs.~12 and 13 in their paper).  On this limited basis, it is difficult to
decide whether the HARP-CDP cross sections are compatible with the dip in the
calculations, although the fact that the HARP-CDP data seem to be consistently
smaller than the HARP data at low momentum is encouraging.

\begin{figure}
  \centering
  \includegraphics[width=\linewidth]{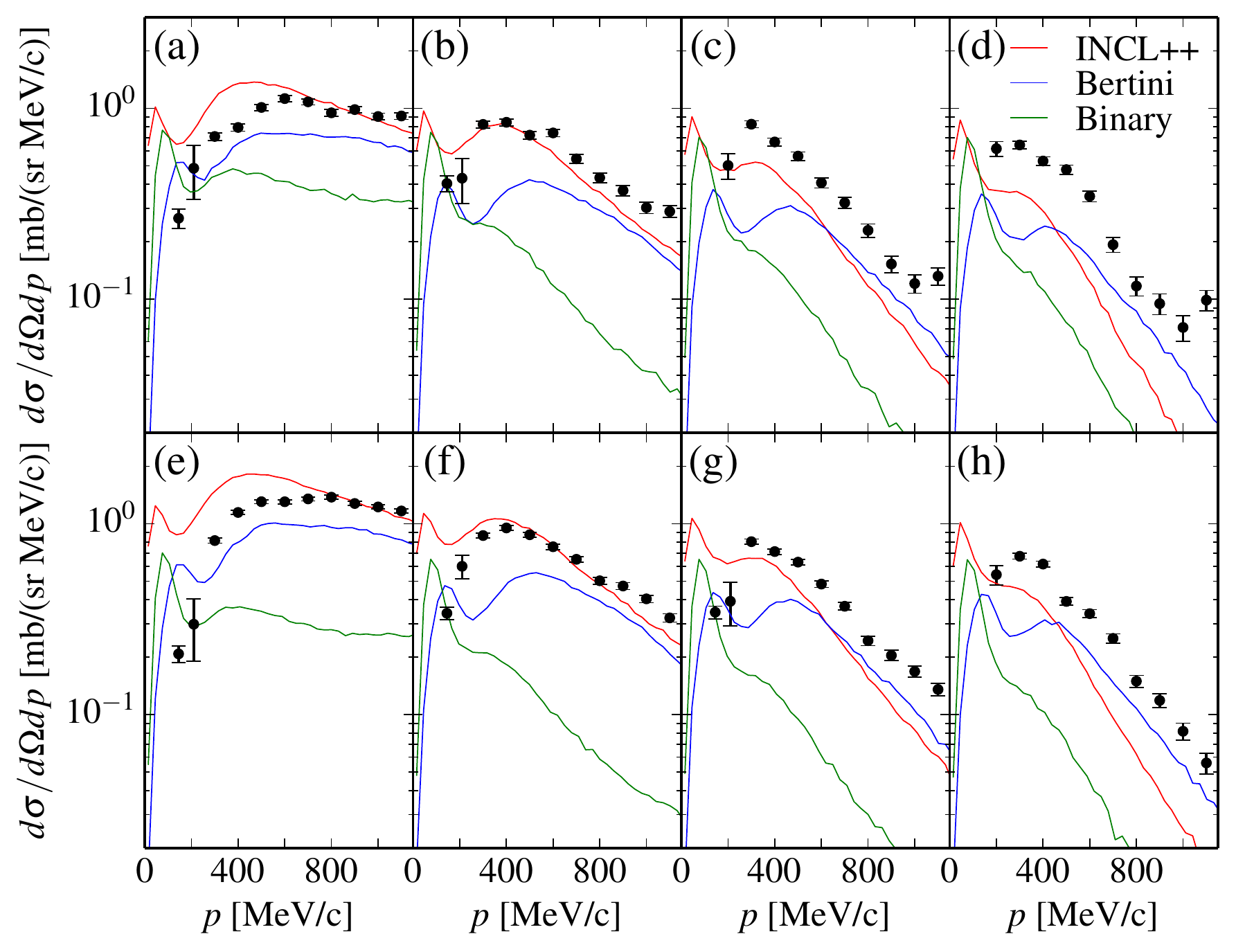}
  \caption{Double-differential cross sections for the production of $\pi^+$ at
    $0^\circ$--$25.8^\circ$ (a, e), $25.8^\circ$--$41.0^\circ$ (b, f),
    $41.0^\circ$--$50.6^\circ$ (c, g) and $50.6^\circ$--$59.0^\circ$ (d, h), from
    12.3~GeV${}/c$ (a--d) and 17.5~GeV${}/c$ (e--h) p+Au. The lines represent
    calculations by different models (see text for details). The experimental
    data are taken from Ref.~\citenum{chemakin-pions}.}
  \label{fig:p_Au_pip_chemakin}
\end{figure}

\begin{figure}
  \centering
  \includegraphics[width=\linewidth]{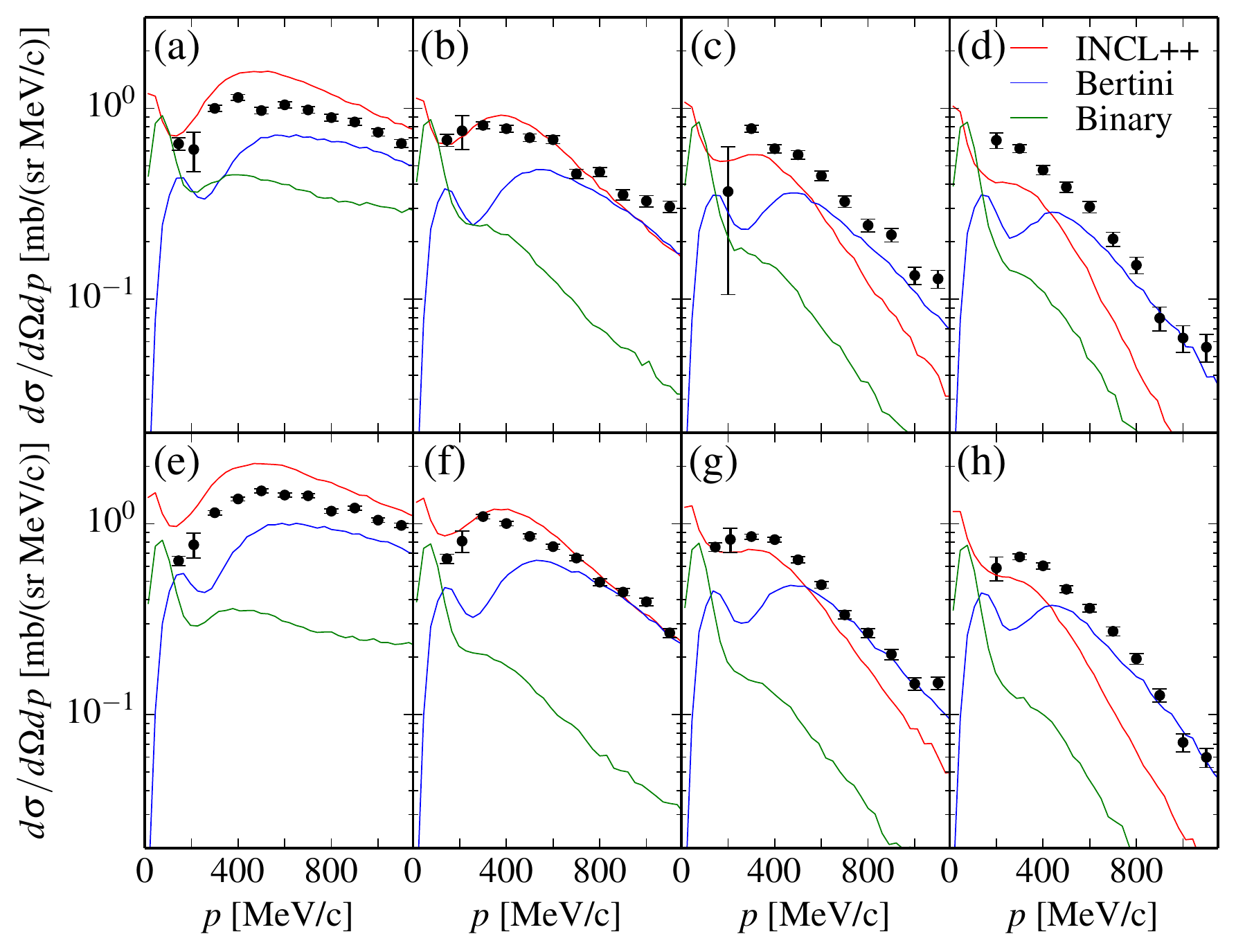}
  \caption{Same as Fig.~\ref{fig:p_Au_pip_chemakin}, but for the production of $\pi^-$.}
  \label{fig:p_Au_pim_chemakin}
\end{figure}

Given this state of affairs, it is surely wise and instructive to consider other
data-sets. Figures~\ref{fig:p_Au_pip_chemakin} and \ref{fig:p_Au_pim_chemakin}
show results for the calculation of double-differential pion-emission cross
sections for $12.3$ and $17.5$ GeV${}/c$ protons on gold targets, compared to the
data from Ref.~\citenum{chemakin-pions}. The energy and the target for the
$12.3$ GeV${}/c$ data-set are close to the HARP data,
Figs.~\ref{fig:p_Pb_compact_ddxs} and \ref{fig:pim_Pb_compact_ddxs}. However,
when comparing the two data-sets, it is important to keep in mind that 1) the
momentum acceptance is larger in Chemakin's data, and 2) the measured angles are
smaller: the largest measurement angle in Chemakin's data-set falls between the
second and the third HARP measurement angle. If one makes abstraction of these
differences, the models appear to behave consistently over all the
Figs.~\ref{fig:p_Pb_compact_ddxs}--\ref{fig:p_Au_pim_chemakin}. Therefore, we do
not see any clear indication that the HARP data should be rejected.

In view of these difficulties, new, high-acceptance data focusing on the pion
production in high-energy proton-induced reaction would be highly
desirable. Together with charged pions, direct measurements of $\pi^{0}$
production would provide fundamental information that could contribute
considerably to the optimization of the INC models.

\section{Thick target: pions in the \ntof{} spallation
  target}\label{sec:thick-targ-second}

\begin{figure}
  \centering
  \includegraphics[width=\linewidth]{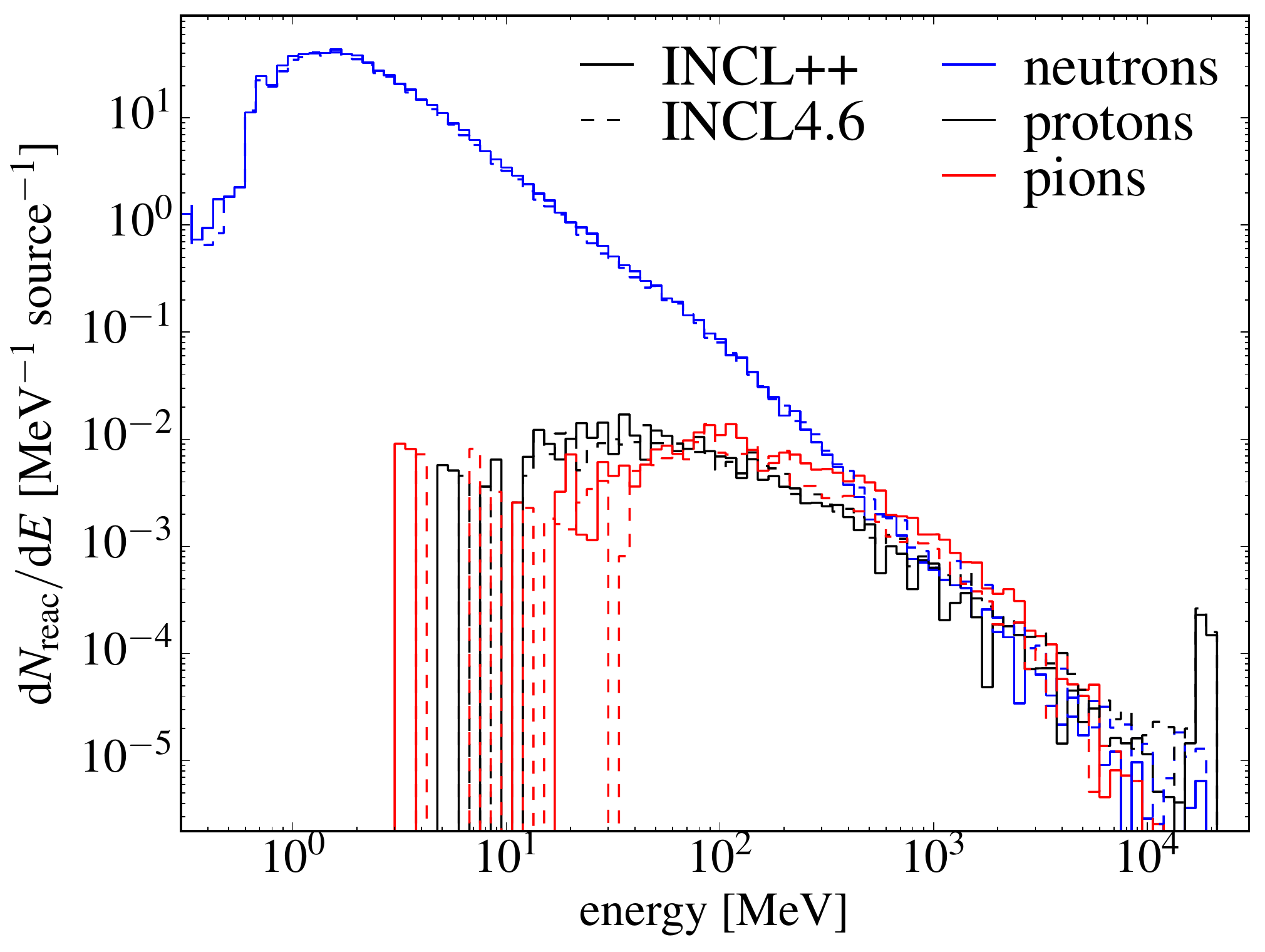}
  \caption{Incident-energy distribution of nuclear reactions induced within the
    hadronic cascade by protons, neutrons and pions, normalized to one primary
    proton, as calculated by the \incl{} model with (\inclxx{}, solid lines) and
    without (\inclf{}, dashed lines) multipion extension, within the \geant{}
    simulation of the \ntof{} spallation target (beam momentum of
    $20$~GeV${}/c$).}
  \label{fig:reaction_distributions}
\end{figure}

While thin-target double-differential cross section can provide some indications
on the ability of the models to correctly predict pion production, the structure
of the hadronic showers that take place in the spallation target, and in
particular the role played by secondary pions in the production of neutrons and
photons, can only be studied by means of dedicated MC simulations of the
spallation process and comparison with available experimental data. In this
respect, we have chosen in this work to perform further simulations of the
\ntof{} spallation target with the \geant{} toolkit. Before analysing the
results of the simulations, it is convenient to briefly describe the toolkit and
the implementation of the MC simulations for the \ntof{} case.

\subsection{The \geant{} toolkit}
\label{sec:geant-toolkit}

\geant{} (GEometry ANd Transport) is a toolkit for the simulation of particle
transport and detector response \cite{agostinelli-geant,allison-geant}.  The
\code{Geant} family of codes was originally developed for the needs of the
high-energy-physics community. However, since the beginning the array of physics
models has been constantly expanding to encompass applications at lower
energy. In particular, \geant{} has been successfully used, since several years,
to describe the transport of neutrons down to thermal energy, using point-wise
cross-section from evaluated libraries
\cite{apostolakis-progress_g4_hadronics}. These developments recently triggered
new work on the use of \geant{} for the simulation of spallation neutron
sources. Recently, \geant{} simulations performed for the \ntof{} source were
benchmarked against experimental results \cite{lo_meo-ntof_g4}, such as the
neutron fluence and resolution function, and yielded interesting results which
will be shortly described in the following subsection.

Physics models in \geant{} are organized and collected in ``physics lists'',
which are specifications of the physical processes (and the associated models)
that should be used in the simulation. The names of the available \geant{}
physics lists are often obtained by concatenating the names of the models used
in the hadronic sector, in decreasing order of incident energy. Thus, for
instance, the \ftfpinclxxhp{} physics list, around which much of the present
work revolves, relies on the Fritiof + pre-equilibrium model (\ftfp{}) at high
energy, the \inclxx{} model at intermediate energies, and the \neutronhp{} model
at low energy.

This work is based on \geant{} \code{v10.1}; however, the \inclxx{} model within
\geant{} was manually upgraded to \code{v5.2.9.4}, which is the version that has
been distributed with \geant{} \code{v10.2} (December 2015).

\subsection{The \ntof{} simulation}
\label{sec:ntof-simulation}

The \ntof{} spallation target is a water-cooled lead cylinder surrounded by an
aluminum container and by a neutron moderator. Its structure is described in
detail in Refs.~\citenum{guerrero-ntof,lo_meo-ntof_g4,lerendegui-ntof_ear2}. A
$20$~GeV${}/c$ proton beam impinges on the base of the lead cylinder at an angle
of approximately 10${}^\circ$. The lead target cylinder can be considered as
thick, in the sense that its size (radius $30$~cm, length $40$~cm) is large
compared to the proton mean free path for inelastic collisions at the beam
energy ($\sim15$~cm). Therefore, the primary proton often triggers a nuclear
reaction inside the Pb target and initiates a hadronic shower which eventually
leads to the production of a large number of particles. A note about our
nomenclature: we refer to all ``non-primary'' particles as \emph{secondaries},
regardless of the reaction generation they appear in, in opposition to the
``primary'' particle incident on the spallation target.

Among the particles escaping from the spallation target, we are particularly
interested in neutrons, which are moderated in water, that can be either normal
or borated, and collimated towards the experimental areas. Simulations of the
\ntof{} spallation target have focused on the reproduction of the measured
energy dependence, resolution function, and spatial distribution of the neutrons
entering the first experimental area (EAR1)
\cite{guerrero-ntof,lo_meo-ntof_g4}. The expected flux in the direction of the
second, new experimental area (EAR2) was also studied in a recent paper
\cite{lerendegui-ntof_ear2}. As shown in
Refs.~\citenum{lo_meo-ntof_g4,lerendegui-ntof_ear2}, these measured quantities
are best reproduced by the simulations using the \ftfpinclxxhp{} and
\qgspinclxxhp{} physics lists. For proton-nucleus reactions, these physics lists
use the Liège Intranuclear Cascade model (\inclxx{}) from $1$~MeV to $20$~GeV
incident energy, and either the Fritiof + pre-equilibrium model (\ftfp{}) or the
Quark-Gluon-String + pre-equilibrium model from $15$~GeV upwards. In the region
where the two model overlap ($15$--$20$~GeV), the choice of the model is
randomly sampled, with linearly-interpolated probabilities between the interval
endpoints (a standard procedure in \geant{}). Since the primary proton beam
energy is $\sim19$~GeV ($20$~GeV${}/c$), it is clear that the \ftfp{} or \qgsp{}
will typically be used \emph{at most} for the simulation of the first inelastic
proton-nucleus reaction\footnote{Except of course for events of specific
  classes, involving small energy losses, like e.g.\ quasi-elastic scattering.};
the rest of the hadronic shower is dominated by the \inclxx{} model. For this
reason, we limit our analysis on the \ftfpinclxxhp{} physics list. The following
section introduces a global description of \inclxx{}, with particular focus on
its pion dynamics.

\subsection{Analysis of secondary reactions}
\label{sec:analys-second-react}

We start by illustrating how nuclear reactions in the target are distributed
with respect to the type of the incident particle and its
energy. Figure~\ref{fig:reaction_distributions} shows the distribution of the
projectile energy for nuclear reactions induced by protons, neutrons and
pions. Each distribution is normalized to the total number of reactions per
primary proton induced by the indicated particle. Note that
Fig.~\ref{fig:reaction_distributions} includes the reactions induced by the
primary protons, which appear as a small peak close to the beam energy, whose
integral roughly amounts to $0.91$~reactions per incident proton; this is
consistent with the thickness of the spallation target, which is of the order of
$2.5$~nuclear mean free paths at the beam energy. Note also that this plot does
\emph{not} include elastic collisions. This choice mainly follows from the
consideration that elastic scattering on lead nuclei does not sensibly modify
the projectile energy for nucleons and pions; in addition, at high energy, the
elastic angular distribution is sensibly forward-peaked and is unlikely to
affect the global flow of energy and momentum within the spallation target.

\begin{table*}
  \centering
  \begin{tabular}{c|ccc|ccc|ccc}
    energy & \multicolumn{3}{c}{\inclxx{}} & \multicolumn{3}{|c}{\inclf{}}&
    \multicolumn{3}{|c}{difference (\%)}\\
     (MeV) & neutrons & protons & pions & neutrons & protons & pions& neutrons & protons & pions\\
    \hline
    $0$--$1$ & $14.4$ & $0$ & $1.88$ & $13.3$ & 0 & $1.40$ & $8.7$ & $0$ & $33.7$ \\
    $1$--$20$ & $157.5$ & $0.08$ & $0.03$ & $144.5$ & $0.08$ & $0.02$ & $9.0$ & $4.3$ & $33.0$\\
    $20$--$200$ & $28.3$ & $1.33$ & $1.59$ & $25.9$ & $1.35$ & $1.26$ & $9.2$ & $-1.5$ & $25.6$\\
    $>200$ & $3.51$ & $2.60$ ($1.67$) & $3.70$ & $4.20$ & $3.07$ ($2.14$) & $2.57$ &
    $-16.5$ & $-15.6$ ($-22.1$) & $44.1$\\
    \hline
    total & $203.7$ & $4.01$ ($3.07$) & $7.20$ & $187.9$ & $4.51$ ($3.56$) & $5.26$ &
    $8.4$ & $-11.0$ ($-13.9$) & $36.8$
  \end{tabular}
  \caption{Average number of reactions per incident proton induced by the
    indicated particles within the specified energy ranges, as calculated by our
    \geant{} simulation with (\inclxx{}) and without (\inclf{}) multipion
    extension. The last three columns show the relative difference between the
    \inclxx{} and \inclf{} calculations. The numbers within brackets in the
    proton columns refer to reactions induced by secondary protons alone.}
  \label{tab:reaction_rates}
\end{table*}

Table~\ref{tab:reaction_rates} presents the integral reaction rates over
selected ranges of incident energy, as calculated by \geant{} simulations using
the \incl{} model with (\inclxx{}) and without (\inclf{}) multipion extension.
The most important feature of Table~\ref{tab:reaction_rates} and
Fig.~\ref{fig:reaction_distributions} is that most of the reactions are actually
neutron-induced reactions at relatively low energy. This is essentially due to
two facts: first, as Fig.~\ref{fig:multiplicities} shows, neutrons are by far
the particles that are most abundantly produced in spallation reactions (note
that the neutron curve has been rescaled by a factor $0.25$). Second, unlike
protons and charged pions, neutrons are not stopped by continuous energy loss,
so their flight through the target is always terminated either by leakage or by
a collision. An additional remark that can be made about the neutron
distribution in Fig.~\ref{fig:reaction_distributions} regards the structures
around $1$~MeV, which correspond to the opening of $(n,n')$ reactions on the
four lead isotopes, while reactions below the thresholds are dominated by
radiative capture on lead isotopes, on the moderating water, as well as on
$(n,\alpha)$ reactions on boron. All these features of neutron transport in
thick spallation targets are well known.

It is instead surprising that secondary reactions induced by pions are twice as
many as those induced by protons. Table~\ref{tab:reaction_rates} shows a
contribution to pion-induced reactions even in the range $0$--$1$~MeV; this
actually corresponds to the absorption of stopped negative pions, whose fate is
to quickly form pionic atoms, decay towards the inner atomic shells and
eventually disappear in a reaction with the nucleus. The lifetime for the whole
process (i.e. formation of the pionic atom, atomic decay and pion absorption) is
of the order of $10^{-10}$~s \cite{schneuwly-pions}, which is much smaller than
the pion intrinsic lifetime ($2.6\cdot10^{-8}$~s). Hence, the decay of stopped
negative pions is actually a rare process.  In some sense this actually makes
negative pions akin to neutrons, insofar as they are bound to induce a nuclear
reaction, or escape from the spallation volume.

Even if one neglects $\pi^-$ absorption at rest, pion-induced reactions are
still more common than proton-induced reactions. This finding clashes with the
typical validation strategy applied to spallation-reaction modeling at lower
energies
\cite{david-intercomparison,leray-intercomparison,intercomparison-website},
which emphasizes the role of proton-induced reactions. In particular, while this
approach is appropriate at ADS energies ($\sim500$--$1000$~MeV), it is clearly
insufficient at higher energies, such as those involved in the \ntof{}
spallation source. Of course the choice of the validation endpoints is also
conditioned by the wide availability of comprehensive data-sets for
proton-induced reactions and the relative scarceness of neutron- and
pion-induced data.

There is one further interesting remark that can be made about
Fig.~\ref{fig:reaction_distributions}, and especially
Table~\ref{tab:reaction_rates}. By comparing the numbers for the two models, one
notices that pions induce about $35\%$ more secondary reactions in \inclxx{}
than in \inclf{}. This is of course a consequence of the fact that the average
pion multiplicity is higher in \inclxx{} than in \inclf{}. It is however
instructive to contrast this $35\%$ difference with the much larger difference
in average multiplicity (Fig.~\ref{fig:multiplicities}), which can be as large
as a factor of $8$ or $9$. If we take the number of pion-induced reactions as a
measure of the number of pions that participate in the hadronic cascade, this
result suggests that the structure of the cascade is relatively insensitive to
the average pion multiplicity predicted by the nuclear-reaction model for the
individual reactions. As we shall see in Sec.~\ref{sec:infl-pion-mult}, this has
important consequences for the sensitivity of secondary-particle yields
(neutrons and photons in particular) to the details of the nuclear-reaction
models.

\subsection{The role of neutral pions}
\label{sec:infl-neutr-pions}

In Ref.~\citenum{lo_meo-ntof_g4}, \geant{} simulations of neutron production
at the \ntof{} spallation target, performed with several physics lists, were
compared between each other and to the energy-differential neutron fluence
measured at \ntof{} in various experimental campaigns. As previously mentioned,
it was shown that the lists based on the \inclxx{} model closely reproduce the
energy dependence of the neutron spectrum and, within a 15--20\% difference, the
integrated yield. All other models are able to reproduce the shape, but
overestimate the yield by a larger factor, of up to 70\%. A different behavior
was instead observed for the production of prompt high-energy photons, i.e.\
those reaching the experimental area within $1$~\textmu{}s of the beam pulse,
and with an energy larger than $10$~MeV. In particular, the most and least
intense photon yields are respectively predicted by the \inclxx{} and \bertini{}
models. In general, an anti-correlation was observed between neutron and
prompt-photon production, with models predicting larger neutron yields
predicting smaller photon yields.  Following this observation, several tests
were performed with \geant{} simulations in order to understand the origin of
the differences among the available intranuclear cascade models for neutron and
prompt-photon production.

\begin{table*}
  \centering
  \begin{tabular}{lrc||c|c|c|c|c|c}
    &&& \code{FTFP\_} & \code{FTFP\_} & \code{QGSP\_} &
    \code{FTFP\_} & \code{QGSP\_} & \code{QGSP\_} \\
    && & \code{INCL46\_HP} & \code{INCLXX\_HP} & \code{INCLXX\_HP} &
    \code{BERT\_HP} & \code{BERT\_HP} & \code{BIC\_HP}\\
    \hline
    (a1)&&number $\pi^0$ & 4.2 & 6.3 & 6.7 & 5.4 & 5.5 & 6.1\\
    (a2)&&number $\pi^\pm$ & 7.5 & 11.0 & 11.5 & 10.2 & 10.0 & 11.5\\
    \hline
    (b1)&\multirow{4}{*}[-1ex]{\rotatebox{90}{photons}}& reference  & 80.7 & 119.0 & 138.0 & 93.0 & 104.4 & 115.4\\
    (b2)&& no $\pi^0$ (\%) & 6.1 & 4.7 & 7.5 & 6.5 & 9.6 & 11.6\\
    (b3)&& no $\pi^\pm$ (\%) & 98.8 & 83.7 & 83.3 & 94.2 & 95.3 & 84.9\\
    (b4)&& $\pi^0\to\pi^\pm$ (\%) & 7.5 & 4.7 & 7.7 & 7.0 & 10.5 & 11.7\\
    \hline
    (c1)&\multirow{4}{*}[-1ex]{\rotatebox{90}{neutrons}}& reference & 454.6 & 513.0 & 420.9 & 658.9 & 575.9 & 531.7\\
    (c2)&& no $\pi^0$ (\%) & 97.4 & 94.5 & 95.2 & 97.1 & 100.4 & 100.8\\
    (c3)&& no $\pi^\pm$ (\%) & 60.1 & 56.2 & 42.9 & 58.1 & 54.1 & 45.3\\
    (c4)&& $\pi^0\to\pi^\pm$ (\%) & 125.5 & 128.0 & 138.3 & 112.6 & 133.2 & 132.4\\
    \hline
    (d1)&    \multirow{4}{*}[-1ex]{\rotatebox{90}{conversion}}
    & $\pi^0\to\gamma$ & 17.9 & 18.0 & 19.0 & 16.1 & 17.3 & 16.7\\
    (d2)&& $\pi^\pm\to\gamma$ & 0.1 & 1.8 & 2.0 & 0.5 & 0.5 & 1.5\\
    (d3)&& $\pi^0\to n$ & 2.8 & 4.5 & 3.0 & 3.5 & -0.5 & -0.7\\
    (d4)&& $\pi^\pm\to n$ & 24.1 & 20.4 & 21.0 & 27.1 & 26.4 & 25.4\\
  \end{tabular}
  \caption{Results of test calculations with suppressed $\pi^0$ decay,
    suppressed $\pi^\pm$ transport, and $\pi^0\to\pi^\pm$ conversion. Columns
    represent different physics lists (\code{FTFP\_INCL46\_HP} being short for
    \ftfpinclxxhp{} with \inclf{}). The meaning of the rows is the following:
    (a1) number of $\pi^0$ produced in the reference calculation; (a2) the same,
    for $\pi^\pm$; (b1--b4) number of produced photons above $10$~MeV (b1) in
    the reference calculation, (b2) without $\pi^0$ decay, (b3) without
    $\pi^\pm$ transport, and (b4) with $\pi^0$ converted to $\pi^\pm$; (c1--c4)
    same four lines for produced neutrons; (d1--d4) conversion ratios from
    $\pi^0$ or $\pi^\pm$ to photons or neutrons. Rows (a1--a2), (b1), (c1) and
    (d1--d4) are averages per incident proton; the statistical uncertainties on
    these values are of the order of a few percent. Rows (b2--b4) and (c2--c4) are
    expressed as percentages of the respective reference result. See text for
    the exact definition of these quantities. The same results are also reported
    in graphical form in Fig.~\ref{fig:pi_kill}.}
  \label{tab:pi_kill}
\end{table*}

\begin{figure}
  \centering
  \includegraphics[width=\linewidth]{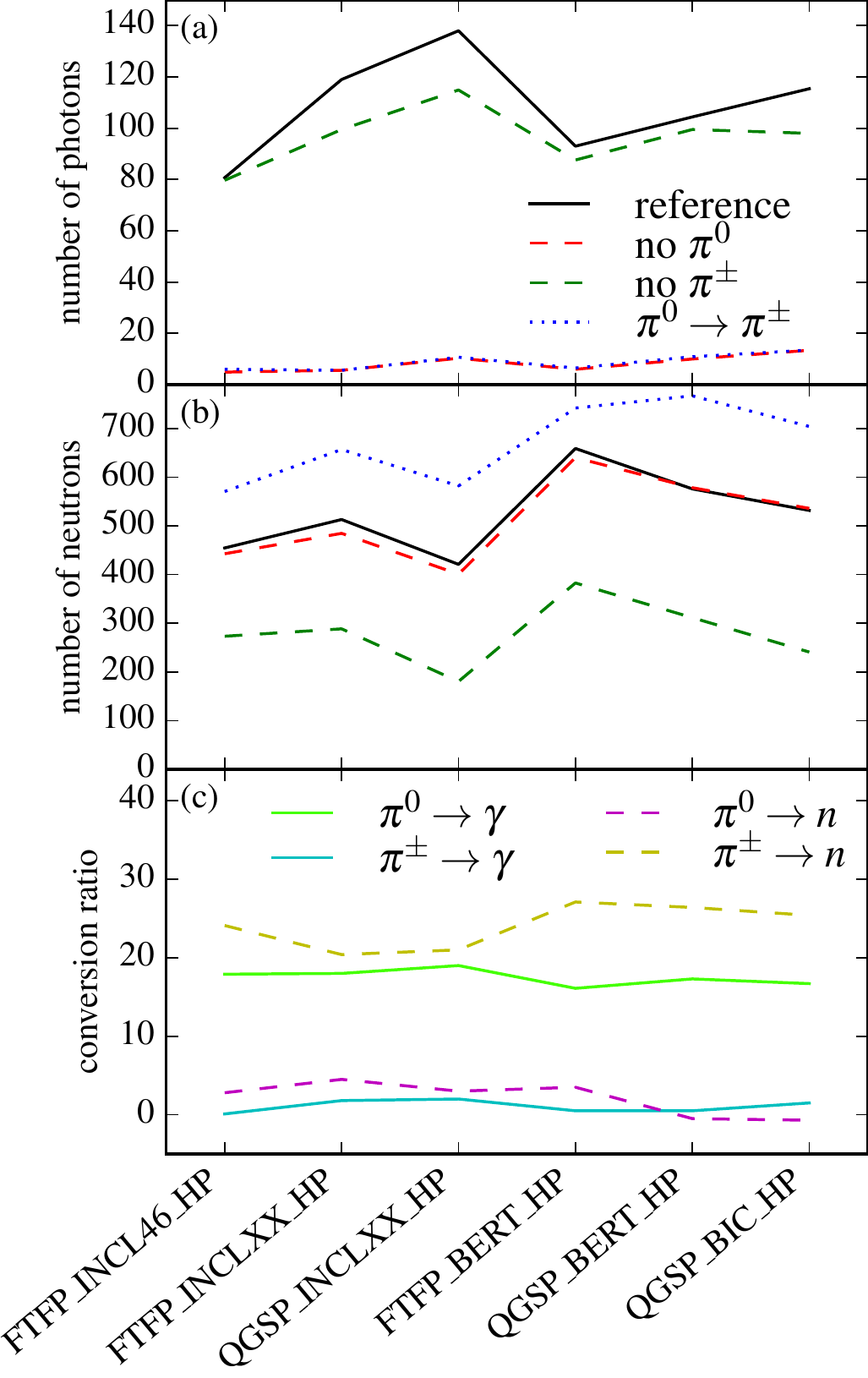}
  \caption{Number of photons (a) and neutrons (b) in the test calculations
    reported in Table~\ref{tab:pi_kill} (rows (b1--b4) and (c1--c4),
    respectively), as a function of the physics list used. Panel (c) represents
    the pion-to-particle conversion ratios (lines (d1--d4) of
    Table~\ref{tab:pi_kill}).}
  \label{fig:pi_kill}
\end{figure}

In order to shed some light on the mechanisms that leads to the production of
photons, we have performed simulations by inhibiting $\pi^0$ decay; this allows
to establish how often prompt photons originate, either directly or through
secondary electromagnetic cascade, from $\pi^0$ decay. A few low-statistics test
runs were performed with different physics lists. The results are shown in
Table~\ref{tab:pi_kill}, which reports the number of produced $\pi^0$s (line
(a1)), along with the number of photons and neutrons produced in the target,
with and without $\pi^0$ decay (lines (b1--b2) and (c1--c2)). All numbers are
normalized to the number of incident protons.  The same results are reported in
graphical form in Fig.~\ref{fig:pi_kill}.

Two minor notes of warning. First, all particles emerging from the reaction
vertex are counted, regardless of the reaction stage that produced them
(intranuclear cascade or statistical de-excitation). Second, the particles are
counted when they are produced, i.e.\ at the reaction site; therefore, in a
sense the same particle can contribute more than once if the reaction cascade it
initiates spans several collisions within the target. Nevertheless, for the sake
of conciseness, we shall make use of expressions such as ``the average number of
photons originating from $\pi^0$ decay''; this is a slight abuse of
terminology. However, for the purpose of ascertaining the influence of pions on
photon and neutron production, this is inconsequential.

Several considerations can be made on the basis of the results in
Table~\ref{tab:pi_kill} and Fig.~\ref{fig:pi_kill}. First, at least $90$\% of
the prompt photons descend in some way from $\pi^0$ decay. This is clearly
indicated by the effect of the suppression of $\pi^0$ decay in all the physics
lists, but also by the observation that physics lists predicting large $\pi^0$
production rates typically also predict larger numbers of photons. The average
$\pi^0$-to-$\gamma$ conversion factor (line (d1)) amounts to about $17$ photons
per $\pi^0$ decay. Second, the influence of $\pi^0$ decay on neutron production
(line (c2)) is sensibly smaller and of the order of the statistical uncertainty
on the reported neutron counts (a few percent). This is easily understood
because the coupling from photon transport to neutron transport is weak; the
$\gamma\to\text{n}$ photonuclear conversion is in general quite inefficient at
producing neutrons, compared to pion-induced reactions.

It is then quite clear that photon yields are positively correlated with (and
actually dominated by) $\pi^0$ production.  However, note also that, in the
reference calculations, large $\pi^0$ production rates (line (a1)) correlate
well with small neutron yields (line (c1)), with \ftfpinclxxhp{} (the physics
list that best reproduces the \ntof{} neutron spectrum) predicting in particular
the highest $\pi^0$ production rate and the lowest neutron yield. This behavior
clashes with the intrinsic pion-to-neutron balance of spallation reactions. In
the framework of INC models, larger values of the pion multiplicity entail more
efficient dissipation of the projectile energy; thus, larger excitation energies
are produced at the end of INC, which results in larger evaporation
yields. Therefore, pion and neutron multiplicities are in principle positively
correlated. The \inclxx{} model can be seen to exhibit this behavior in
Fig.~\ref{fig:multiplicities}: the multipion extension has indeed the effect of
increasing \emph{all} particle multiplicities.  However, while
Fig.~\ref{fig:multiplicities} refers to a thin target case, for a thick target
other effects come into play. As discussed later on in this paper, due to the
resilience of the hadronic cascade, neutron production in large-volume
spallation targets is rather insensitive to the pion multiplicity in individual
$N+N$ or $\pi+N$ collisions, but depends mostly on the total number of pions
produced in the full hadronic cascade. In particular, for a larger number of
pions produced, 40\% of which are $\pi^{0}$, a larger fraction of energy will be
diverted from the hadronic cascade to the electromagnetic one, and become
unavailable for neutron production. In this sense, the results in
Table~\ref{tab:pi_kill} and Fig.~\ref{fig:pi_kill} clearly hint to a fundamental
role of pion production in general, and $\pi^{0}$ production in particular, in
determining the final neutron and photon fluences in spallation neutron sources
based on high-energy protons impinging on large spallation volumes.

In summary, it is quite clear that neutral pions dominate the production of
prompt, high-energy photons. At the same time, the observed anti-correlation of
neutron and prompt-photon yields can be explained considering that neutral pions
essentially divert energy from the hadronic cascade towards the electromagnetic
one.

\subsection{Influence of charged pions on neutron production}
\label{sec:infl-charg-pions}

As we have shown in Fig.~\ref{fig:reaction_distributions} and
Table~\ref{tab:reaction_rates} above, charged pions are responsible for a large
number of secondary reactions. Since spallation reactions often lead to the
production of neutrons, one expects that secondary charged pions have a large
influence on neutron production.

In order to prove this, we have performed test calculations in which we suppress
the transport of charged pions by killing their tracks at the reaction
site. This prevents them from inducing further secondary
reactions. Table~\ref{tab:pi_kill} and Fig.~\ref{fig:pi_kill} show the number of
produced $\pi^\pm$ tracks (line (a2)), along with the average photon (line (b3))
and neutron yields (line (c3)) recorded when charged pions are not
transported. Pion-induced reactions are seen to account for about $40$--$45$\%
of the neutron yield. Depending on the physics list, the suppression of one
charged pion entails a reduction of $20$--$25$ neutrons per incident proton
(line (d4)), with \incl{}-based and \bertini{}-based physics lists respectively
yielding the lowest ($20.4$) and highest ($27.1$) pion-to-neutron conversion
ratios. There is also an effect on the photon yields (line (b3)), which also
decrease by $5$--$20$\% when charged pions are suppressed. This is probably due
to the disappearance of the neutral pions that would have been produced in the
suppressed $\pi^\pm$-induced reactions.

To further substantiate our claims about the importance of secondary pions, we
have performed test calculations where $\pi^0$ production is randomly replaced
with the production of a $\pi^+$ or a $\pi^-$ (with equal probability) with the
same kinetic energy and momentum direction as the suppressed $\pi^0$. In this
test, energy is essentially conserved (up to the small $\pi^0/\pi^\pm$ mass
difference), and charge is conserved on average. The effect on photon yields
(line (b4)) is essentially the same as in the calculation without $\pi^0$ decay,
which is reasonable; neutron yields (line (c4)), on the other hand, are
increased by about $10$--$30$\%, depending on the physics lists. Note that the
increased neutron yields are consistent with the conversion efficiency that can
be estimated from the $\pi^0\to n$ and $\pi^\pm\to n$ conversion coefficients
(lines (d3--d4)) and the number of transformed $\pi^0$ (line (a1)).

\subsection{Influence of the physics list}
\label{sec:infl-phys-list}

It is worth spending a few words about the dependence of the numbers in
Table~\ref{tab:pi_kill} on the choice of the physics list. In particular, it is
instructive to compare results obtained with the same INC/de-excitation
component, but different high-energy reaction models. When doing so, care must
be exercised because the switching energy between parton-string models and INC
depends on the physics list. For instance, in the case of neutron-nucleus
reactions, the INC stage is used up to $9.9$~GeV in \code{QGSP\_BERT\_HP}, up to
$5$~GeV in \code{FTFP\_BERT\_HP} and up to $20$~GeV in \qgspinclxxhp{} and
\ftfpinclxxhp{}.

Therefore, we focus on the most straightforward comparison, namely the one
between \qgspinclxxhp{} and \ftfpinclxxhp{}, which use the same fade-out energy
interval between \inclxx{} and the parton-string model
($15$--$20$~GeV). Table~\ref{tab:pi_kill} shows that \qgspinclxxhp{} produces
$16$\% larger photon yields and $18$\% smaller neutron yields. This is
qualitatively consistent with the observation by \citet{lerendegui-ntof_ear2}
that \qgspinclxxhp{} yields smaller neutron fluences; however, the numbers in
Table~\ref{tab:pi_kill} are not directly comparable with
\citeauthor{lerendegui-ntof_ear2}'s integrated neutron fluences, because our
results include counts of unobserved in-target tracks.

The comparison between \code{QGSP\_BERT\_HP} and \code{FTFP\_BERT\_HP} leads to
the same kind of conclusions, albeit quantitative comparisons are
difficult. Inspection of the differences between \ftfpinclxxhp{} and
\code{FTFP\_BERT\_HP}, for instance, suggests that the influence of the
INC/de-excitation stage is of comparable magnitude.

\subsection{Influence of pion multiplicity}
\label{sec:infl-pion-mult}

The tests involving the suppression of pion emission might be considered as
unrealistic because they grossly violate several conservation laws. Indeed, each
pion carries several hundreds MeV of mass and kinetic energy, which can be
(partially) converted into neutrons (through nuclear reactions) or photons
(through nuclear reactions or $\pi^0$ decay) if the pion is not
suppressed. Charge conservation is also violated when charged pions are
killed. The $\pi^0\to\pi^\pm$ transformation test, on the other hand,
approximately conserves energy (up to the $\pi^0/\pi^\pm$ mass difference),
respects charge conservation on average, and provides insight to the sensitivity
of the simulation to the pion charge distribution.

It is also instructive to study the dependence of the calculation results on the
mean pion multiplicity in individual nuclear reactions. One way to test the
influence of multiplicity without abandoning energy conservation is to compare
the results of calculations performed with nuclear-reaction models that yield
different average pion multiplicities. It is however difficult to isolate the
effect of the change in pion multiplicity from all the other differences between
the models. Ideally, one would like to change the average pion multiplicity and
nothing else. The closest approach to an ideal setting for a sensitivity study
is to modify the nuclear-reaction physics within the same model. By acting on
some internal model parameter, one could modify the average pion multiplicity
while keeping the overall coherence of the model. 
%
For this purpose, we shall discuss calculations performed with \inclxx{} with
and without multipion extension (see Sec.~\ref{sec:model-descr-liege}). 
We remind the reader that pion multiplicities are much smaller in the model
without multipion extension, starting from about $1$~GeV incident energy
(Fig.~\ref{fig:multiplicities}).

\begin{figure}
  \centering
  \includegraphics[width=\linewidth]{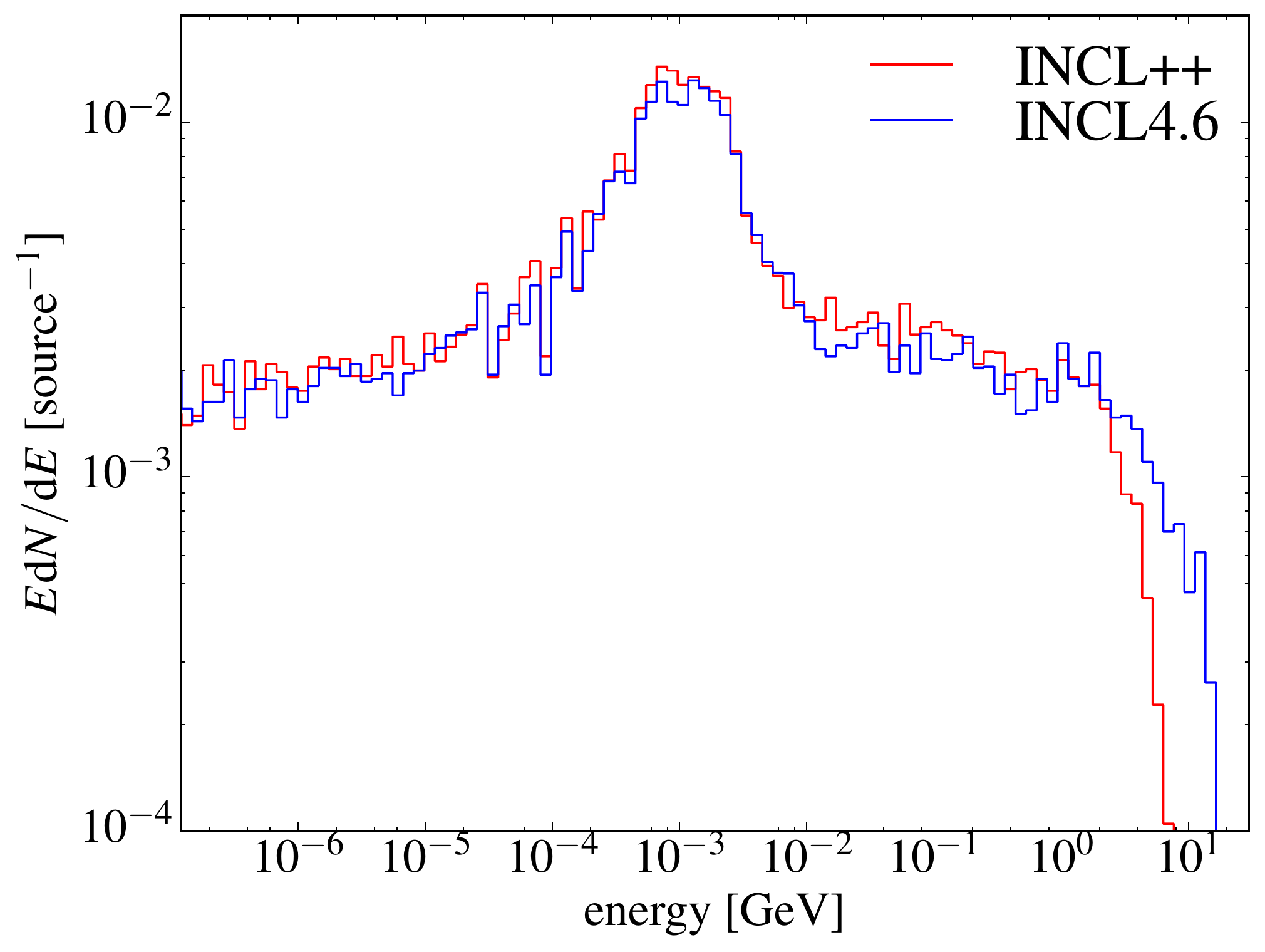}
  \caption{Neutron spectrum per unit lethargy emitted from the \ntof{}
    spallation source in a $2^\circ$ cone directed towards the first
    experimental area, as calculated by our \geant{} simulation. Results are
    shown for the \incl{} model with (\inclxx) and without (\inclf) multipion
    extension.}
  \label{fig:neutrons_incl46_inclxx}
\end{figure}

\begin{figure}
  \centering
  \includegraphics[width=\linewidth]{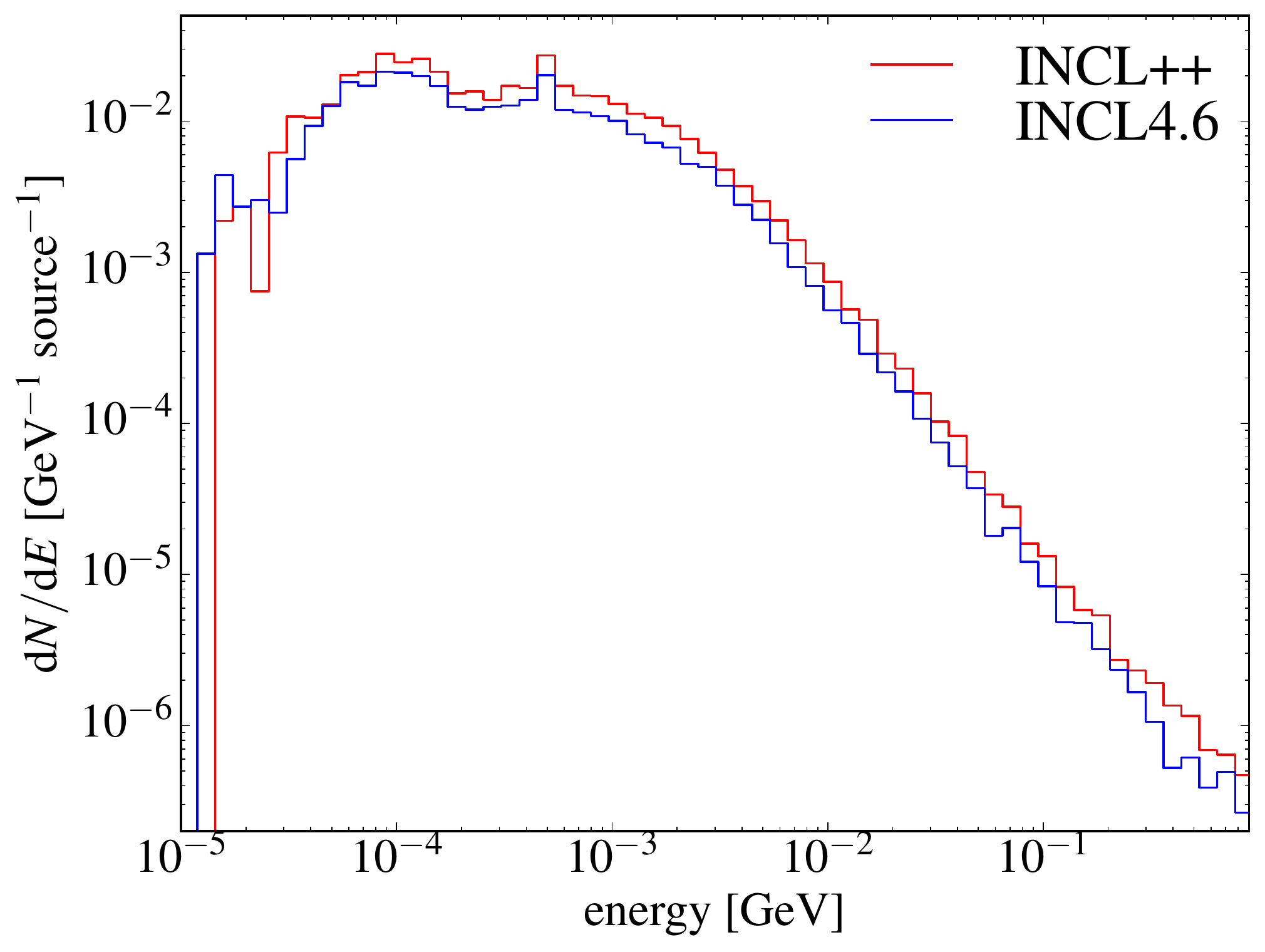}
  \caption{Photon spectrum emitted from the spallation source within $100$~ns of
    the beam pulse, in a $2^\circ$ cone directed towards the first experimental
    area, as calculated by our \geant{} simulation. Results are shown for the
    \incl{} model with (\inclxx) and without (\inclf) multipion extension.}
  \label{fig:photons_incl46_inclxx}
\end{figure}

\begin{table}
  \centering
  \begin{tabular}{c|c|c|c}
    energy & \inclxx{} & \inclf{} & difference \\
    (GeV) & ($\times10^{-3}$) & ($\times10^{-3}$) & (\%) \\
    \hline
    $0$--$1$ & $72.1$ &  $66.9$ & $+7.7$\\
    $1$--$20$ & $2.4$ &  $3.5$ & $-31.4$ \\
    \hline
    total & $74.5$  & $70.4$ & $+5.8$
  \end{tabular}
  \caption{Average number of neutrons per incident proton emitted in a $2^\circ$
    cone directed towards the first experimental area, in the indicated energy
    ranges, as calculated by our \geant{} simulation. Results are shown for the
    \incl{} model with (\inclxx) and without (\inclf) multipion extension.}
  \label{tab:neutrons_incl46_inclxx}
\end{table}

\begin{table}
  \centering
  \begin{tabular}{c|c|c|c}
    energy & \inclxx{} & \inclf{} & difference \\
    (MeV) & ($\times10^{-3}$) & ($\times10^{-3}$) & (\%) \\
    \hline
    $0$--$10$ & $50.9$ & $37.8$  & $+34.7$\\
    $>10$ & $11.6$  & $8.0$  & $+45.0$ \\
    \hline
    total & $62.5$  & $45.8$ & $+36.5$
  \end{tabular}
  \caption{Same as Table~\ref{tab:neutrons_incl46_inclxx}, for the emission of
    photons within $100$~ns of the beam pulse.}
  \label{tab:photons_incl46_inclxx}
\end{table}

Figure~\ref{fig:neutrons_incl46_inclxx} shows the calculated neutron fluence per
unit lethargy and per incident proton emitted from the spallation source in the
direction of the first experimental area. This spectrum is scored at the exit of
our geometry and we only consider neutrons within a $2^\circ$ cone around the
direction towards EAR1. This spectrum actually represents the input for the
resampling procedure that leads to the estimation of the fluence at EAR1, which
is 185~m away \cite{lo_meo-ntof_g4}.

It is immediately obvious that the difference between the two models is much
smaller than the differences seen in the previous
section. Table~\ref{tab:neutrons_incl46_inclxx} shows the average numbers of
neutrons produced in selected energy ranges. The global effect of the multipion
extension on the neutron yield is of the order of a few percent. This should be
compared to the reduction of $\sim40$\% that was induced by the suppression of
the charged-pion tracks.

The neutron spectra are essentially identical in shape up to about
$\sim1$~GeV. As expected, the shape of the low-energy ($E\lesssim20$~MeV) end of
the spectrum, which is treated using evaluated cross-section databases, is
independent of the high-energy model chain that feeds it. This remark has
already been made in connection with the comparison of the spectra predicted by
different physics lists \cite{lo_meo-ntof_g4}. In the case of the comparison
between \inclf{} and \inclxx{}, the similarity stretches up to $1$~GeV because
the two models indeed yield very similar predictions up to this energy (cfr.\
Fig.~\ref{fig:multiplicities}). Neutrons above $1$~GeV are seen to be more
abundantly produced by \inclf{}. This is coherent with the aforementioned
observation (Sec.~\ref{sec:infl-neutr-pions}) that higher pion multiplicities
lead to more efficient dissipation of the projectile energy; clearly, it is
easier to produce high-energy neutrons if less energy is channeled into pion
production. In this respect, the validity of the multipion extension could be
verified by examining the neutron fluence produced at the \ntof{} spallation
source in the energy region above 1~GeV. However, as of now there is no
experimental information available, and in any case the measurement of the
neutron fluence at such high energy is not straightforward.

Figure~\ref{fig:photons_incl46_inclxx} and Table~\ref{tab:photons_incl46_inclxx}
compare the results of \inclf- and \inclxx-based calculations on prompt-photon
production. We apply the same $2^\circ$ angular cut on the angle of the emitted
photon; in addition, we only select photons that are emitted within $100$~ns
from the beam pulse. This roughly corresponds to the prompt photons that are
detected before $1$~\textmu{}s at the EAR1 experimental area
\cite{lo_meo-ntof_g4}.

The difference between the calculations is much larger than for neutrons,
especially for the production of high-energy photons. This is not surprising,
given the crucial role that neutral pions are seen to play in the production of
prompt high-energy photons. However, compared to the results of
Sec.~\ref{sec:infl-neutr-pions} above, the magnitude of the effect is somewhat
mitigated.

\begin{figure} 
  \centering
  \includegraphics[width=\linewidth]{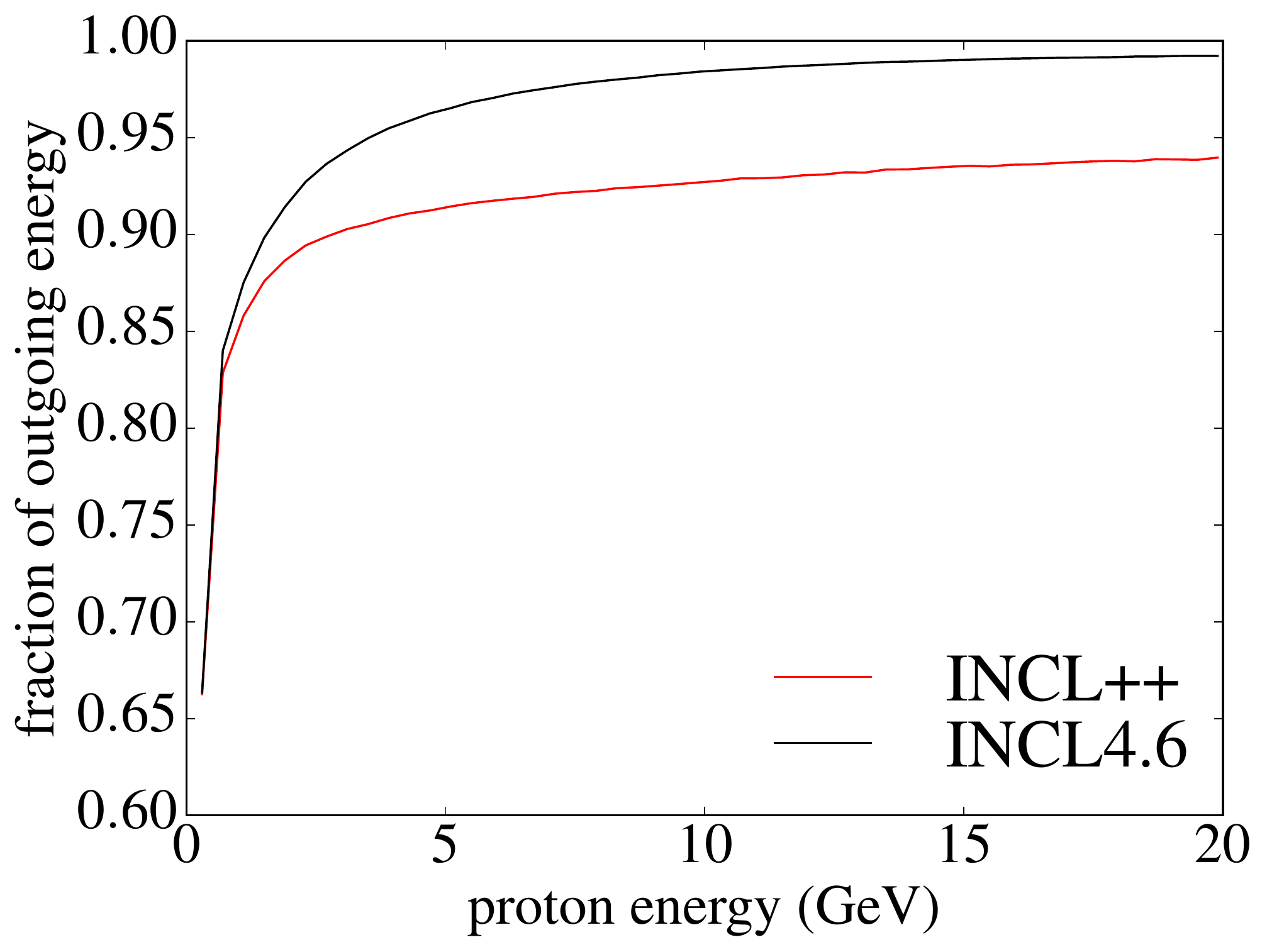}
  \caption{Excitation function for the average fraction of outgoing energy in
    p+$^{208}$Pb reactions, as calculated with (\inclxx{}) and without
    (\inclf{}) multipion extension. The masses of escaping $\pi^\pm$ are counted
    as outgoing energy. Neutral pions and photons are not counted.}
  \label{fig:outgoing_energy_incl46_inclxx}
\end{figure}


The results shown above indicate that the pion multiplicity in a single
nucleon-nucleon or pion-nucleon collision does not play a crucial role in
neutron and photon production in thick spallation targets, contrary to the large
effect predicted in the thin target case and shown in
Fig.~\ref{fig:multiplicities}. This property has already been observed at lower
energies \cite{aoust-pion_physics}; our work confirms that it still holds at
\ntof{} energies. The fact that photon and neutron production are not affected
by the reduced pion multiplicity of \inclf{} (Sec.~\ref{sec:infl-pion-mult}) can
be partly ascribed to the fact that in this model pions are still produced,
albeit not as abundantly as in \inclxx{}. However, a more general argument to
explain the relative insensitivity of the photon and neutron yields to the
details of the INC model involves a rather peculiar feature of thick spallation
targets, which can actually be inferred from most thick spallation-target
benchmarks \cite[see e.g.][]{koi-satif12_benchmark}. In thick spallation
targets, in fact, it is the structure of the hadronic cascade that plays a
compensatory role.

To better describe the effect, it is also useful to look at the outgoing energy
of a reaction (in this case p+$^{208}$Pb), defined as the sum of all the kinetic
energies of all the reaction products, except photons and $\pi^0$s. For outgoing
$\pi^\pm$, their mass is also counted as outgoing energy. The outgoing energy
represents the maximum energy which is liable to be injected in further nuclear
reactions (neglecting photonuclear
reactions). Figure~\ref{fig:outgoing_energy_incl46_inclxx} shows the ratio
between the outgoing energy and the incident projectile energy, as a function of
the projectile energy, in p+$^{208}$Pb, as calculated with \inclf{} and
\inclxx{}. The two models yield very similar predictions, even at $20$~GeV, with
the difference never exceeding 5\% of the incident energy. This similarity
should be contrasted with Fig.~\ref{fig:multiplicities}, which demonstrates that
the events generated by the two models are radically different.

Of course, the insensitivity of the outgoing energy to the details of the model
is partly a trivial consequence of the fact that energy and momentum are
conserved in both cases. Yet, the curves in
Figure~\ref{fig:outgoing_energy_incl46_inclxx} have important consequences for
the development of the hadronic cascade in a thick target. For the sake of
clarity we assume here that secondary particles have a mean free path for
inelastic collisions which is short with respect to the size of the geometry
and, in the case of charged particles, to the range. If this assumption holds,
as it may often be the case for thick spallation targets, particles leaving the
first reaction vertex are likely to induce secondary reactions nearby, leading
to further degradation of the projectile energy. Therefore, for a less
dissipative model (such as \inclf) it will take a greater number of soft
reactions to degrade the projectile energy into low-energy particles. On the
contrary, a more dissipative model (such as \inclxx) will require fewer hard
reactions. However, on a geometric scale larger than the reaction mean free
path, the overall structure of the resulting hadronic cascade will be relatively
independent of the details of the model. In other words, the number of neutrons
and photons produced in the full hadronic cascade (and therefore the fluence of
particles emerging from the spallation target) will mostly depend on the total
number of pions (and other intermediate particles) produced in the cascade,
rather than on their multiplicity in each reaction.

It remains to be seen if the conditions for a short mean free path are met in
the case of the \ntof{} spallation target. The reaction cross sections for
high-energy protons, neutrons and pions in lead are weakly dependent on energy
and all of the order of $1.8$~b, which results in mean free paths of the order
of $15$~cm. This is sensibly smaller than the thickness of the spallation target
($40$~cm). A $15$-cm range in lead corresponds to a proton energy of about
$400$~MeV, or a pion energy of $250$~MeV. Therefore, for particle energies above
roughly $600$~MeV (and for neutrons of all energies), the aforementioned
conditions are approximately met and the target may be considered as thick.


\section{Conclusions}
\label{sec:conclusions}

Triggered from recent works on \geant{} simulations of the \ntof{} spallation
target, which indicated some sizable differences in neutron production between
various INC models, we have here investigated the role of secondary pions in
high-energy proton-induced reactions, both for thin and thick targets. A
comparison with the available experimental data points to a large
underestimation of pion production by the previous version of the Li\`ege
intranuclear-cascade model. Some shortcoming is also observed for the Binary and
Bertini code in reproducing measured double-differential cross sections for
charged pion production in p+Pb reactions at high energy ($12$~GeV/$c$). On the
contrary, the recent version of the Li\`ege Intranuclear Cascade Model,
\inclxx{}, with multipion extension, reproduce measured cross sections
reasonably well at all energies, except for the presence of a dip in the
momentum distribution at forward angles (predicted also by the Bertini
code). The presence of this dip seems to be related to formation and decay of
the $\Delta(1232)$ resonance. However, the scarce data available do not allow at
present to draw a conclusion on this effect. Fresh new data in this respect
would be highly desirable, together with measurements of neutral pion
production, which are essentially missing up to date.

The role of pion production in spallation reactions has been further
investigated by performing dedicated \geant{} simulations of the \ntof{}
spallation target. As suggested in Ref.~\citenum{lo_meo-ntof_g4}, models
producing overall more pions per incident proton also produce fewer neutrons,
with a clear anti-correlation effect. On the contrary, prompt $\gamma$-rays,
which are mostly produced by electromagnetic cascade following $\pi^{0}$ decay,
correlate with the total number of pions in the hadronic cascade. This
observation points out to a fundamental role of $\pi^{0}$ production in
determining both the final neutron and $\gamma$-ray fluence produced by a
spallation source based on high-energy protons on a heavy target, a phenomenon
which is well known in the context of calorimetry detection in high-energy
physics \cite{fabjan-calorimetry_review}.

It has been further observed that, contrary to the thin-target case, the
previous \incl{} version leads to thick-target neutron and photon productions
that are not sensibly different from the new version, except for very
high-energy neutrons (above a few GeV). This finding has been related to the
structure of the whole hadronic cascade that develops in a thick spallation
target; the emitted particles are shown to be sensitive only to the total number
of pions produced along the hadronic cascade, in particular neutral pions. The
mitigated dependence of thick-target yields on the underlying elementary cross
sections is a direct consequence of conservation laws, and applies only for
dimensions of the spallation target sensibly larger than the mean free path of
secondary particles, as in the case of the \ntof{} target considered in this
work. This is valuable insight about the remarkable resilience of hadronic
cascades and their manifest insensitivity to the details of the underlying
physical models.

It remains to discuss to which extent secondary pions can be held responsible
for the differences in neutron and photon yields predicted by the different
\geant{} physics lists \cite{lo_meo-ntof_g4}. While our work shows that
pion-induced reactions can be relatively important, it should be borne in mind
that other secondary particles are also at play. Secondary proton-nucleus
reactions are seen to be less abundant than pion-nucleus reactions at these
energies (see Fig.~\ref{fig:reaction_distributions} and
Table~\ref{tab:reaction_rates}) and are probably better constrained in the
models, due to the wide availability of elementary experimental data. This is
much less the case for secondary neutron-nucleus reactions, which are sparsely
represented in the elementary benchmarks of spallation models
\cite[e.g.][]{david-intercomparison,leray-intercomparison,intercomparison-website},
despite the fact that they are very common
(Fig.~\ref{fig:reaction_distributions} and Table~\ref{tab:reaction_rates}). We
believe that differences in the treatment of secondary neutrons and pions can
account for most of the variation among the \geant{} physics lists.

\begin{acknowledgement}
  This work was partially financed by the CHANDA EU FP7 project (grant agreement
  605203).
\end{acknowledgement}


\end{document}